\documentclass[11pt]{article}

\usepackage{hyperref} 
\hypersetup{colorlinks=true, linkcolor=blue, citecolor=blue, urlcolor=blue}

\usepackage[linesnumbered, ruled, vlined]{algorithm2e} 
\usepackage{amsmath} 
\usepackage{amssymb} 
\usepackage{amsthm} 
\usepackage{bm, bbm} 
\usepackage{booktabs} 
\usepackage{cleveref} 
\usepackage{enumitem} 
\usepackage{ifthen} 
\usepackage{lipsum} 
\usepackage{makecell} 
\usepackage{mathrsfs} 
\usepackage{mathtools} 
\usepackage{multirow} 
\usepackage[sort,numbers]{natbib}
\usepackage{optidef} 
\usepackage{orcidlink} 
\usepackage{physics} 
\usepackage{tikz} 
\usepackage{xparse} 
\usepackage{siunitx} 

\usepackage[a4paper,margin=30mm]{geometry}

\usepackage{CJK}

\definecolor{cA}{HTML}{0072BD}
\definecolor{cB}{HTML}{EDB120}
\definecolor{cC}{HTML}{77AC30}
\definecolor{cD}{HTML}{D95319}

\newcommand{\Cpp}{C
\nolinebreak[4]
\hspace{-.05em}\raisebox{.4ex}{\relsize{-3}{\textbf{++}}}}
\newcommand{\defeq}{\coloneqq}

\newcommand{\relmiddle}[1]{\mathrel{}\middle#1\mathrel{}}

\ExplSyntaxOn
\NewDocumentCommand{\definealphabet}{mmmm}{ \int_step_inline:nnn{`#3}{`#4}{ \cs_new_protected:cpx{#1 \char_generate:nn{##1}{11}}{ \exp_not:N #2{\char_generate:nn{##1}{11}}}}}
\ExplSyntaxOff

\definealphabet{bb}{\mathbb}{A}{Z}
\definealphabet{rm}{\mathrm}{A}{Z}
\definealphabet{cal}{\mathcal}{A}{Z}
\definealphabet{frak}{\mathfrak}{a}{z}

\crefname{equation}{Eq.}{Eqs.}
\crefname{figure}{Fig.}{Figs.}
\crefname{tabular}{Tab.}{Tabs.}
\crefname{section}{Sec.}{Secs.}
\crefname{subsection}{Sec.}{Secs.}
\crefname{algorithm}{Algorithm}{Algorithms}

\allowdisplaybreaks[4]

\usetikzlibrary{
  3d,
  calc,
  math,
  matrix,
  patterns,
  backgrounds,
  arrows.meta,
  shapes.geometric,
  decorations.pathmorphing,
}

\newboolean{isMain}
\setboolean{isMain}{true}

\begin{document}
\title{ Initial Placement for Fruchterman--Reingold Force Model with 
  Coordinate Newton Direction }
\author{ Hiroki Hamaguchi\,\orcidlink{0009-0005-7348-1356} Naoki Marumo\,\orcidlink{0000-0002-7372-4275}
  Akiko Takeda\,\orcidlink{0000-0002-8846-4496}
}
\maketitle

\begin{abstract}
  The Fruchterman--Reingold (FR) force model is widely used in force-directed graph drawing, and multilevel approaches such as \textsf{sfdp} in Graphviz scale these methods effectively.
  A crucial step in multilevel schemes is refinement, which improves the graph layout, typically performed with the simulation-based algorithm.
  For this refinement, we can utilize optimization-based methods such as L-BFGS, or combine them with initial placement methods such as Simulated Annealing to achieve better layouts.
  However, they have several limitations, such as suffering from high per-iteration costs for large graphs or having difficulty with weighted and structurally complex graphs, leaving room for improvement.

  In this research, we propose a new initial placement based on stochastic coordinate descent to accelerate the optimization process.
  We first reformulate the problem as a discrete optimization problem using a hexagonal lattice and then iteratively update a randomly selected vertex along the coordinate Newton direction with low per-iteration costs.
  We demonstrate the effectiveness of our method through numerical experiments, showing that our initial placement leads to faster convergence and higher-quality layouts compared to naive optimization approaches.
  We also discuss applications of our method, such as drawing for Hooke--Coulomb and Eades force models.
\end{abstract}

\section{Introduction}
\label{sec:introduction}

\begin{figure}[t]
  \centering
  \includegraphics[width=0.8\columnwidth]{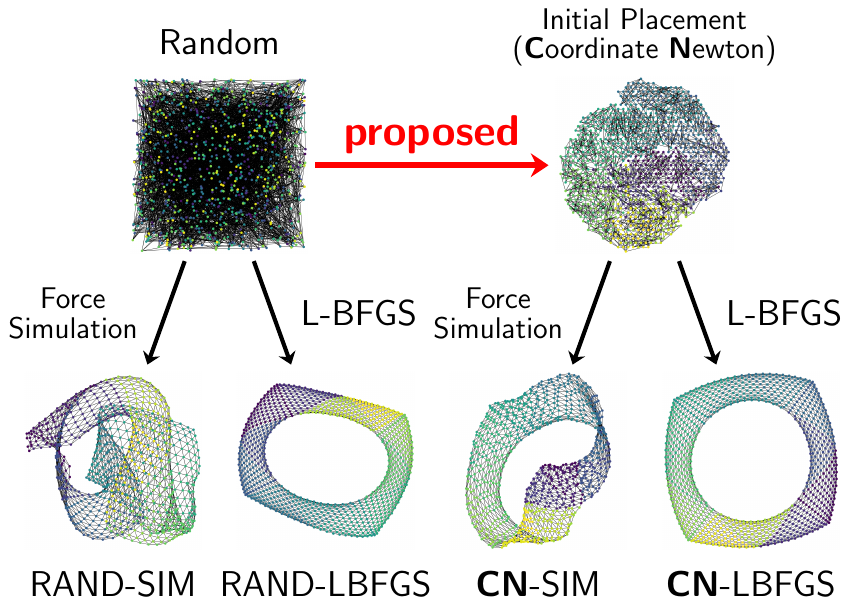}
  \caption{Comparison of the algorithms for the \texttt{jagmesh1}.}
  \label{fig:fig1}
\end{figure}

Graph drawing is a fundamental task in information visualization, and force-directed methods are among the most widely used approaches.
In these methods, vertices are treated as particles subject to forces, and the resulting equilibrium layouts are regarded as desirable visualizations.
Among established force models~\citep{eades1984heuristic,kamadaAlgorithmDrawingGeneral1989}, the Fruchterman--Reingold (FR) force model~\citep{fruchtermanGraphDrawingForcedirected1991} is particularly prominent for its intuitive formulation, flexibility, and simplicity.
It is implemented in libraries such as NetworkX~\citep{hagbergExploringNetworkStructure2008}, Graphviz~\citep{ellsonGraphvizOpenSource2002}, and igraph~\citep{csardiIgraphSoftwarePackage2006}.

A central challenge in force-directed graph drawing is efficiently obtaining an equilibrium state of the force system.
In force-directed methods, the refinement process iteratively adjusts vertex positions to reduce force imbalance, and its efficiency strongly affects both the computational cost and the quality of the visualization.
This refinement remains important even in multilevel approaches such as the Scalable Force-Directed Placement (\textsf{sfdp}) algorithm~\citep{Hu2006EfficientHF,ellsonGraphvizOpenSource2002}, where layouts must still be refined while propagating solutions from coarse to fine graph levels.

Two major approaches are commonly used for refinement.
The first is a simulation-based approach \citep{fruchtermanGraphDrawingForcedirected1991}, which repeatedly applies the forces to vertices to approximate equilibrium.
The second is numerical optimization based on explicit energy minimization.
In particular, the L-BFGS (Limited-memory Broyden--Fletcher--Goldfarb--Shanno) algorithm~\citep{liuLimitedMemoryBFGS1989a}, a general-purpose optimization method, has been reported to produce high-quality layouts for several force models~\citep{6183577}, including the FR force model and Kamada--Kawai (KK) model~\citep{kamadaAlgorithmDrawingGeneral1989}.
Although they are effective, both approaches require many computationally expensive iterations.
When the number of iterations is insufficient, the resulting layouts may contain twists, including folded or tangled structures that reduce readability~\cite{veldhuizenDynamicMultilevelGraph2007,cheongSnapshotVisualizationComplex2018}, as shown in \cref{fig:fig1}.

One possible approach to mitigating the issue is to improve the initial placement before refinement.
Since both the simulation-based algorithm and the L-BFGS algorithm involve relatively expensive iterations, it is often more efficient to first move the vertices toward a reasonably good configuration using inexpensive approximate updates, and then apply full-cost refinement methods only after the layout becomes sufficiently close to equilibrium.
This pre-processing substitutes for part of the refinement while reducing the overall optimization difficulty.

A pre-processing step based on Simulated Annealing (SA) has been shown to improve the final visualization quality when combined with the simulation-based algorithm~\cite{ghassemitoosiSimulatedAnnealingPreProcessing2016}.
However, the existing method assumes circular initial layouts and employs a relatively heuristic objective function, insufficiently grounded in the underlying force model.
In addition, the method mainly targets unweighted, small-scale, and structurally simple graphs, limiting its applicability to more general graph drawing settings.

In this paper, we propose a new initial placement to quickly obtain the equilibrium layout of the force models, especially the FR force model.
We provide an initial placement with fewer twists than random placement within a short time, accelerating the subsequent optimization process.
We can use both the simulation-based and L-BFGS algorithms to obtain the final placement.
This work extends the applicability of the initial placement idea to larger-scale, weighted, and complicated structured graphs.
To achieve this, we optimize the position of vertices one by one with the coordinate Newton direction, leveraging the inherent structure and the sparsity of graphs.
We also demonstrate its effectiveness through various experiments and discuss its potential applications, such as drawing for the HC (Hooke--Coulomb) \citep{atallahAlgorithmsTheoryComputation2009} and Eades \citep{eades1984heuristic} force models.

The remainder of the paper is organized as follows.
\begin{enumerate}[label=\textbullet, itemsep=0pt, parsep=0pt]
  \item In \cref{sec:preliminaries}, we provide the technical definitions of the problem.

  \item In \cref{sec:related_work}, we review some related work.

  \item In \cref{sec:algorithm}, we present our initial placement algorithm.

  \item In \cref{sec:experiment}, we report experimental comparisons.

  \item In \cref{sec:discussion}, we discuss the potential applications and limitations.

  \item In \cref{sec:conclusion}, we conclude the paper and discuss future work.
\end{enumerate}

\section{Preliminaries}
\label{sec:preliminaries}

\begin{figure}[t]
  \centering
  \includegraphics[width=0.6\columnwidth]{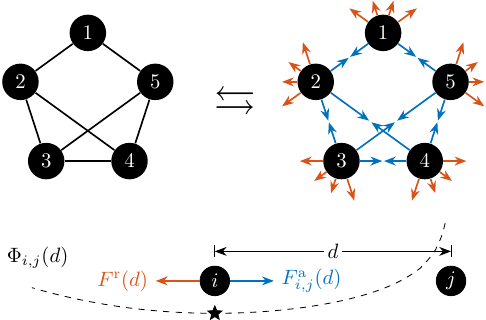}
  \caption{ In the model, forces act on every pair of vertices.
    Forces $F_{i,j}^{\mathrm{a}}(d)$ and $F^{\mathrm{r}}(d)$ work between vertices $i$ and $j$.
    The equilibrium of them is achieved at $d = k/\sqrt[3]{a_{i,j}}$, which equals $k$ when $a_{i,j}= 1$.
  }
  \label{fig:frLayout}
\end{figure}

In this study, we adopt an energy-based optimization approach for graph drawing.
While Fruchterman and Reingold~\citep{fruchtermanGraphDrawingForcedirected1991} originally proposed the force-directed simulation-based algorithm based on a physical particle system, energy-based methods achieve the same goal by directly minimizing an energy function $\Phi$.
These two perspectives are fundamentally equivalent because a local minimum of the energy function $\Phi(X)$ yields an equilibrium position since the negative gradient $-\nabla \Phi(X) = 0$ corresponds to the forces.
In this sense, the force and energy perspectives are equivalent as both capture equilibrium, illustrated in \cref{fig:frLayout}.

Let us formally define the energy minimization problem.
Let $G = (V, E)$ be an undirected graph with vertex set $V = \{1, \dots, n\}$ and edge set $E$.
Each edge $(i,j) \in E$ $(i<j)$ has positive weight $a_{i,j} = a_{j,i} \in \mathbb{R}$.
For convenience, we set $a_{i,j}=0$ for $(i, j) \notin E$.
In this problem, the force model assumes forces between vertices.
For vertices $i$ and $j$ with distance $d > 0$, an attractive force $F_{i,j}^{\mathrm{a}}(d)$ and a repulsive force $F^{\mathrm{r}}(d)$ are defined as
\begin{equation*}
  F_{i,j}^{\mathrm{a}}(d) \coloneqq \frac{a_{i,j}d^{2}}{k}, \quad F^{\mathrm{r}}
  (d) \coloneqq -\frac{k^{2}}{d},
\end{equation*}
where $k > 0$ is a constant parameter, often set to $1/\sqrt{n}$~\citep{fruchtermanGraphDrawingForcedirected1991}.
The scalar potentials (or energies) of these forces~\citep{6183577} are given by
\begin{equation*}
  \Phi^{\mathrm{a}}_{i,j}(d) \coloneqq \int_{0}^{d}F_{i,j}^{\mathrm{a}}(r) \dd{r}= \frac{a_{i,j}d^{3}}{3k}, \quad
  \Phi^{\mathrm{r}}(d) \coloneqq \int_{1}^{d}F^{\mathrm{r}} (r) \dd{r}= -k^{2}\log{d}.
\end{equation*}
For simplicity, we define $\Phi^{\mathrm{r}}(0)=\infty$.%
\footnote{We can also use $-k^{2}\log(d + \epsilon^{\mathrm{r}})$ to prevent divergence, where $\epsilon^{\mathrm{r}}$ is constant.}
Let $\norm{\cdot}$ denote the Euclidean norm in $\mathbb{R}^{2}$.
Then, the problem is to minimize the energy with $X \coloneqq (x_{1}, \dots, x_{n}) \in \mathbb{R}^{2 \times n}$:
\begin{mini}
  {X \in \mathbb{R}^{2 \times n}} {\Phi(X) \coloneqq \sum_{(i,j) \in E} \Phi_{i,j}^{\mathrm{a}}(\norm{x_i - x_j}) + \sum_{i<j} \Phi^{\mathrm{r}}(\norm{x_i - x_j}).}
  {\label{prob:fr}} {}
\end{mini}
We refer to Problem~\ref{prob:fr} as the energy minimization problem for the FR force model.

\begin{figure}[t]
  \centering
  \includegraphics[width=0.6\columnwidth]{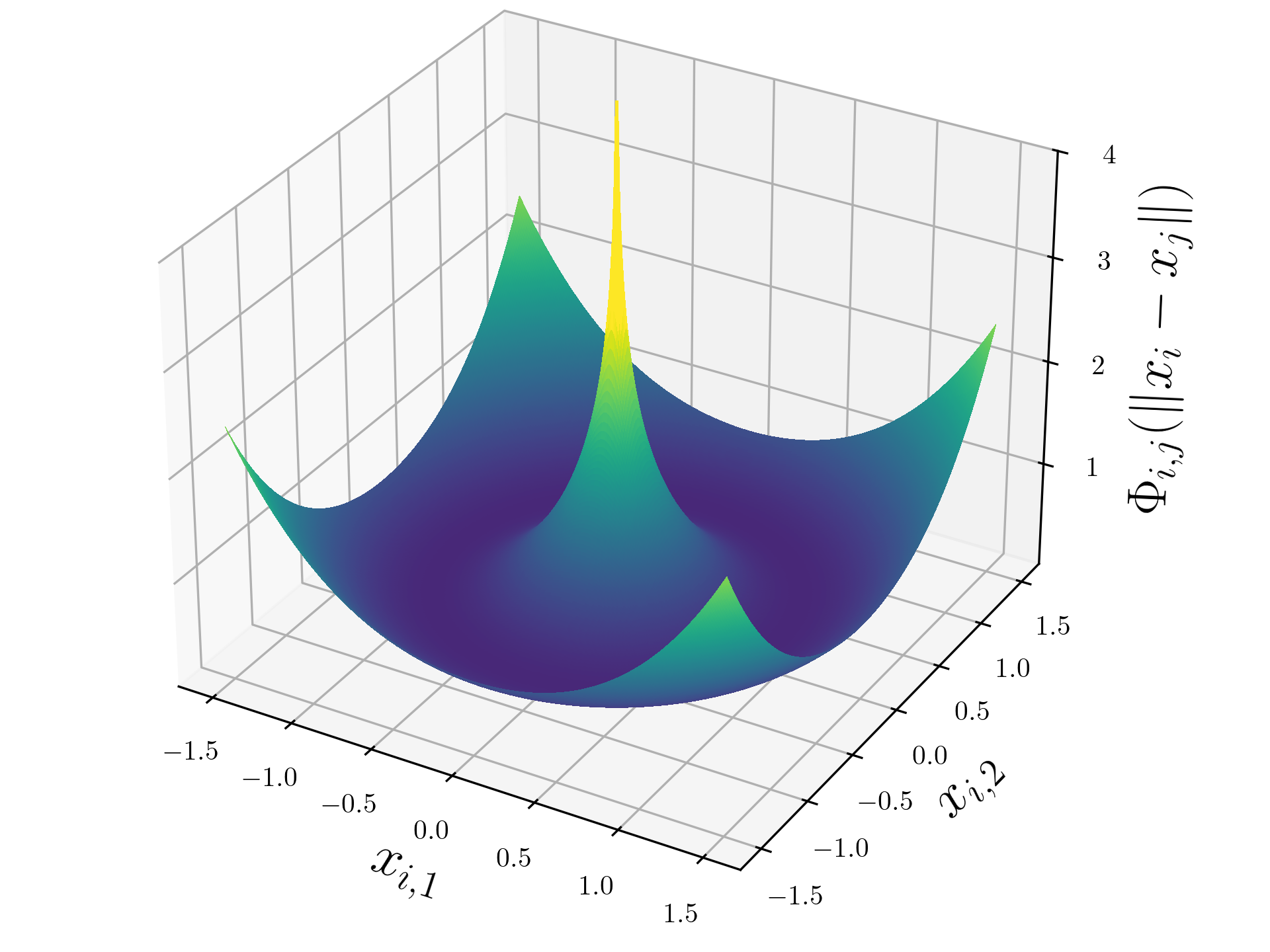}
  \caption{The energy $\Phi_{i,j}(\norm{x_i - x_j})$ for $x_{i}=(x_{i,1},x_{i,2})$, $x_{j}=(0,0), a_{i,j}= 1$, and $k = 1$.}
  \label{fig:energy3d}
\end{figure}

We also define the vertex pairwise energy function $\Phi_{i,j}(d) \coloneqq \Phi_{i,j}^{\mathrm{a}}(d) + \Phi^{\mathrm{r}}(d)$ for convenience.
Here, $\Phi_{i,j}(\cdot)=\Phi_{j,i}(\cdot)$ holds for all $i,j\in V$.
While the energy function $\Phi_{i,j}(d)$ is convex and minimized when $d = k/\sqrt[3]{a_{i,j}}$ for a positive $a_{i,j}$, a function $x_{i}\mapsto \Phi_{i,j} (\norm{x_i - x_j})$ is non-convex for a fixed $x_{j}$.
Additionally, $\Phi_{i,j}$ is not Lipschitz continuous near $d=0$, as illustrated in \cref{fig:energy3d}.
These properties highlight the difficulty of the optimization problem.

\section{Related Work}
\label{sec:related_work}

We next summarize the existing literature relevant to our work.

\subsection{Optimization Approach}
\label{sec:optimization_approach}

Optimization-based methods for graph drawing have been actively investigated.
Representative approaches include GPU acceleration~\citep{gajdosParallelFruchtermanReingold2016}, machine learning techniques~\citep{tiezziGraphNeuralNetworks2024}, multi-objective formulations~\citep{ahmedMulticriteriaScalableGraph2022a}, and stochastic gradient descent~\citep{zhengGraphDrawingStochastic2019}.
Among these, our attention is on the underlying optimization algorithms themselves.

The energy minimization problem can be interpreted as a continuous optimization problem.
Let $\Phi_i$ denote the contribution of vertex $x_i$ to the objective function $\Phi$, defined by
\begin{equation*}
  \Phi_{i}(x_{i}) \coloneqq \sum_{j \in V \setminus \{i\}} \Phi_{i,j}(\norm{x_{i}- x_{j}}).
\end{equation*}
The $i$-th column of the gradient $\nabla \Phi(X)\in\mathbb{R}^{2\times n}$ is given by
\begin{equation*}
  (\nabla \Phi(X))_{i} \defeq \nabla \Phi_{i}(x_{i}) = \sum_{j \in V \setminus \{i\}}\left(\frac{a_{i,j}\norm{ x_{i}- x_{j}}}{k}- \frac{k^{2}}{\norm{ x_{i}- x_{j}}^{2}}\right) (x_{i}- x_{j}).
\end{equation*}
Since both the objective value and gradient are explicitly available, a variety of gradient-based optimization methods can be applied to solve the energy minimization problem.
The classical simulation-based algorithm (also known as the FR algorithm, presented in \cref{alg:fr}) may be interpreted as a steepest-descent-type method in which the step size is controlled by a cooling schedule.
Indeed, $-(\nabla \Phi(X))_i$ corresponds to the force acting on vertex $i$.
The L-BFGS algorithm, a renowned and widely used optimization method, is another first-order gradient-based method that has been shown to consistently outperform the original simulation-based algorithm in speed and layout quality~\citep{6183577}.
A common choice for the initial solution is a random placement of nodes within the bounding box~\citep{2020SciPy-NMeth}.

However, when the input is highly twisted or otherwise ill-conditioned, both the simulation-based algorithm and the L-BFGS algorithm can suffer from high computational costs. Each iteration operates on the full set of coordinates, which limits its ability to untangle local geometric defects, and the total number of iterations required to reach a good layout can be large.
Thus, for pre-processing purposes, it is desirable to use inexpensive updates that move the layout toward a reasonably good configuration before applying the simulation-based algorithm or L-BFGS.
This study therefore proposes stochastic coordinate descent with coordinate Newton directions as a lightweight initial-placement method.

\subsection{Initial Placement Approach}
\label{ssec:preprocessing}

A preprocessing step with Simulated Annealing (SA) is known to be effective~\citep{ghassemitoosiSimulatedAnnealingPreProcessing2016} since SA can avoid getting stuck in local optima and leads to better visualization, combined with the simulation-based algorithm.
Let
\begin{equation*}
  Q^{\mathrm{circle}} \defeq \left\{\qty(\cos(\frac{2\pi k}{n}), \sin(\frac{2\pi k}{n})) \relmiddle| 1 \leq k \leq n \right\}
\end{equation*}
be the points on a unit circle in $\bbR^{2}$.
For an unweighted graph $G$, let $E_{2}$ be a set of vertex pairs with a shortest path distance equal to two.
Let $\angle(a, b)$ denote the angle between the lines from the origin to the points $a$ and $b$, measured in the interval $(-\pi, \pi]$.
This study~\citep{ghassemitoosiSimulatedAnnealingPreProcessing2016} defines the problem for the preprocessing of graph drawing as follows:
\begin{mini}
  {X \in \bbR^{2 \times n}} {\sum_{(i,j)\in E \cup E_2} \abs{\angle(x_i, x_j)},}
  {\label{eq:sa}} {} \addConstraint{x_i}{\in Q^\mathrm{circle} \quad}{\text{for $1 \leq i \leq n$}}
  \addConstraint{x_i}{\neq x_j \quad}{\text{for $1 \leq i < j \leq n$}.}
\end{mini}
Problem~\eqref{eq:sa} is a discrete optimization problem where the placement is limited to $Q^{\mathrm{circle}}$ and uses angles, not the function $\Phi$ in the energy minimization problem.
This study obtains faster and better visualization by setting the result of SA for Problem~\eqref{eq:sa} as an initial placement for the simulation-based algorithm.

Still, the following limitations remain:
\begin{enumerate}[label=\textbullet, itemsep=0pt, parsep=0pt]
  \item The target graphs are restricted to unweighted ones.

  \item The layout is confined to a simple circle, which could be ineffective for complex structured graphs.

  \item The neighborhood in the SA is the random swapping of two vertices, making the optimization process inefficient for large-scale graphs.

  \item $\abs{E_2}$ could be $\Theta(\abs{V}^{2})$, unable to leverage the sparsity of graphs if it exists.
\end{enumerate}
We address these limitations by proposing a new initial placement algorithm.

\section{Proposed Algorithm}
\label{sec:algorithm}

With the prior studies in \cref{sec:optimization_approach}, we propose a new initial placement algorithm for the energy minimization problem.
The algorithm solves a discrete optimization problem with stochastic coordinate descent to find an initial placement with fewer twists.
This algorithm leverages the properties of the objective function to determine the initial layout quickly.
By combining it with the simulation-based algorithm and the L-BFGS algorithm, it enhances the overall convergence speed of graph drawing through optimization.

\subsection{Newton Direction and Coordinate Newton Direction}
\label{ssec:introNewton}

Let us first explain the Newton direction and the coordinate Newton direction, the key concept in our algorithm.
Consider a strictly convex function $g \colon \bbR^{n}\to \bbR$ at $x_{0}$.
The Newton direction $d = -\nabla^{2}g(x_{0})^{-1}\nabla g(x_{0})$ is an optimal direction for the second-order approximation of $g$:
\begin{equation*}
  g(x_{0}) + \nabla g(x_{0})^{\top}(x - x_{0}) + \frac{1}{2}(x - x_{0})^{\top}\nabla
  ^{2}g(x_{0}) (x - x_{0}),
\end{equation*}
where $x = x_{0}+ d$ is the minimizer of this approximation.
Note that the Hessian matrix $\nabla^{2}g(x_{0})$ is positive definite since $g$ is strictly convex.
Although the Newton direction is essential in various iterative methods, it requires the computation of the inverse Hessian $\nabla^{2}g(x_{0})^{-1}\in \bbR^{n \times n}$, posing a high computational cost for large-scale problems.

Still, we can leverage the concept of the Newton direction in a different manner: the coordinate Newton direction.
Instead of computing the inverse Hessian $\nabla^{2}g(x_{0})^{-1}$ in the entire variable space $\bbR^{n}$, we restrict the variable $x$ to its coordinate block $x_{i}$ with fewer dimensions, and compute $\nabla^{2}g_{i}(x_{i})^{-1}\nabla g_{i}(x_{i})$ where $g_{i}$ is a restricted function of $g$ to $x_{i}$.
Since the coordinate Newton direction computation is much cheaper than that of the Newton direction, we can repeat this procedure many times.
In general, this idea is known as stochastic coordinate descent~\citep{wright2022optimization} or Randomized Subspace Newton~\citep{3454287.3454343} in a broader context.

In particular, this coordinate Newton direction has an apparent natural affinity to the energy minimization problem.
We can compute the coordinate Newton direction by taking the position $x_{i}$ of the vertex $i$ as the coordinate block.
Although directly applying this idea to the energy minimization problem is challenging, as we will discuss in \cref{sec:rationale}, we leverage this coordinate Newton direction to propose our algorithm.

\subsection{Discrete Optimization Problem for Initial Placement}
\label{ssec:reduction}

Even at the expense of accuracy, obtaining an approximate solution quickly is crucial for the initial placement.
To obtain it, we simplify the energy minimization problem into a more manageable and well-behaved discrete optimization problem:
\begin{mini}
  {X \in \bbR^{2 \times n}} {\Phi^\mathrm{a}(X) \defeq \sum_{(i,j)\in E} \frac{a_{i,j}\norm{x_i - x_j}^{3}}{3k},}
  {\label{eq:frApprox0}} {} \addConstraint{x_i}{\in Q^\mathrm{hex} \quad}{\text{for $1 \leq i \leq n$}}
  \addConstraint{x_i}{\neq x_j \quad}{\text{for $1 \leq i < j \leq n$},}
\end{mini}
where
\begin{equation}
  \label{eq:qHex}Q^{\mathrm{hex}}\defeq \left\{ \qty(q+\frac{1}{2}r, \frac{\sqrt{3}}{2}
  r) \relmiddle| q \in \bbZ,\ r \in \bbZ \right\}.
\end{equation}

\begin{figure}[t]
  \centering
  \includegraphics[width=0.7\columnwidth]{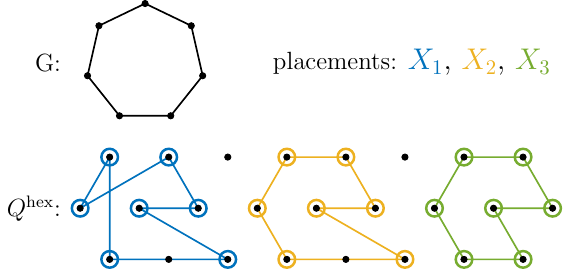}
  \caption{ Concept of $Q$.
    The assignment from $V$ to a discrete point placement $Q$, especially a hexagonal lattice $Q^{\mathrm{hex}}$.
    Apparently, among the three placements, the right one is the best placement for Problem~\eqref{eq:frApprox0}.
  }
  \label{fig:pi}
\end{figure}

Let us explain the rationale for this formulation.
First, recall that the energy minimization problem is equivalent to the following:
\begin{mini*}
  {X \in \bbR^{2 \times n}} {\sum_{(i,j)\in E} \frac{a_{i,j}\norm{x_i - x_j}^{3}}{3k}- \sum_{i<j} k^2\log{\norm{x_i - x_j}}.}
  {} {}
\end{mini*}
This formulation separates the $\order{\abs{E}}$ terms from $\Phi^{\mathrm{a}}_{i,j}$ and the $\order*{\abs{V}^2}$ terms from $\Phi^{\mathrm{r}}$.
Following previous research mentioned in \cref{ssec:preprocessing}, we fix the possible positions $x_{i}$ of each vertex $i$ to a finite discrete set of points $Q$, where $\abs{Q}\geq \abs{V}$.
We also impose a condition on $Q$ such that $\norm{q_{i}- q_{j}}\geq 1$ for all $q_{i}, q_{j}\in Q$ with $q_{i}\neq q_{j}$.
This makes the second $\order*{\abs{V}^2}$ terms from $\Phi^{\mathrm{r}}$ bounded by a constant.
Since $\Phi^{\mathrm{r}}(d)=-k^{2}\log{d}$ is a convex function such that it decreases monotonically with respect to $d$, for sufficiently large $d$, the value of $-k^{2}\log{d}$ does not vary excessively.
We can also prevent the divergence of the energy function by the condition $\norm{q_{i}- q_{j}}\geq 1$.
Under these observations, we drop the second term and derive the problem as:
\begin{mini*}
  {X \in \bbR^{2 \times n}} {\sum_{(i,j)\in E} \frac{a_{i,j}\norm{x_i - x_j}^{3}}{3k},}
  {} {} \addConstraint{x_i}{\in Q \quad}{\text{for $1 \leq i \leq n$}}
  \addConstraint{x_i}{\neq x_j \quad}{\text{for $1 \leq i < j \leq n$}.}
\end{mini*}
This means that we skip to consider the $\order*{\abs{V}^2}$ pairs by fixing the possible point placement in advance, reducing the computational complexity to $\order{\abs{E}}$ and thus offering significant speedup.
See \cref{fig:pi} for a visual explanation.

In this study, we adopt a hexagonal lattice $Q^{\mathrm{hex}}$~\citep{patelHexagonalGrids2013,s22145179} defined by~\cref{eq:qHex} (see \cref{fig:pi}) as $Q$.
When minimizing the objective function that arises from the attractive energy $\Phi^{\mathrm{a}}$, it is advantageous for the points to cluster as closely as possible.
In this context, the hexagonal lattice is known for its densest packing structure in space \citep{changSimpleProofThues2010} and offers computational simplicity.

We note that the scaling of the lattice structure does not affect the optimization process.
As we will see later, our algorithm is scale invariant, meaning that the distance between $\norm{q_{i}- q_{j}}$ does not affect the update.

\subsection{Newton Direction for Discrete Optimization}
\label{ssec:newtonDirection}

Next, we formulate an optimization procedure for Problem~\eqref{eq:frApprox0}.
Although it is challenging to solve, the coordinate Newton direction of a randomly selected vertex $i$ provides significant insights as mentioned in \cref{ssec:introNewton}.
Let the restricted objective function $\Phi^{\mathrm{a}}_{i}(x_{i})$ be
\begin{equation*}
  \Phi^{\mathrm{a}}_{i}(x_{i}) \defeq \sum_{j \in V \setminus \{i\}}\frac{a_{i,j}\norm{x_i - x_j}^{3}}{3k}.
\end{equation*}
Its gradient and Hessian matrix are
\begin{align*}
  \nabla \Phi^{\mathrm{a}}_{i}(x_{i})    & = \sum_{j \in V \setminus \{i\}}\frac{a_{i,j}\norm{x_i - x_j}}{k}(x_{i}- x_{j}),                             \\
  \nabla^{2}\Phi^{\mathrm{a}}_{i}(x_{i}) & = \sum_{j \in V \setminus \{i\}}\frac{a_{i,j}\norm{x_i - x_j}}{k}\mqty(1                                 & 0 \\
  0                                      & 1) + \sum_{j \in V \setminus \{i\}}\frac{a_{i,j}}{k\norm{x_i - x_j}}(x_{i}- x_{j})(x_{i}- x_{j})^{\top}.
\end{align*}
Note that $\Phi^{\mathrm{a}}_{i}$ is strictly convex, assuring the Hessian matrix $\nabla^{2}\Phi^{\mathrm{a}}_{i}(x_{i})$ is positive definite.

The ordinary updated rule with the coordinate Newton direction is
\begin{equation*}
  x_{i}^{\mathrm{new}}\gets x_{i}- \nabla^{2}\Phi^{\mathrm{a}}_{i}(x_{i})^{-1}\nabla
  \Phi^{\mathrm{a}}_{i}(x_{i}).
\end{equation*}
Note that when we scale the lattice structure by a factor of $s > 0$, the gradient is scaled by $s^{2}$ and the Hessian matrix is scaled by $s$, resulting in the coordinate Newton direction being scaled by $s$.
Thus, the algorithm for the FR force model is scale invariant.
Since $x_{i}^{\mathrm{new}}$ may not be in the hexagonal lattice $Q^{\mathrm{hex}}$ in Problem~\eqref{eq:frApprox0}, we project this $x_{i}^{\mathrm{new}}$ onto the nearest point in $Q^{\mathrm{hex}}$.
We also empirically found that adding a random noise vector to the direction is effective for the optimization process, a strategy similar to the SA in \cref{ssec:preprocessing}.
This randomness can help to escape from local minima and to explore the solution space more effectively.
In conclusion, the update rule for the vertex $i$ is
\begin{equation*}
  x_{i}^{\mathrm{new}}\gets \mathrm{round}\qty(x_{i}- \nabla^{2} \Phi^{\mathrm{a}}_{i}
  (x_{i})^{-1}\nabla \Phi^{\mathrm{a}}_{i}(x_{i}) + r),
\end{equation*}
where $\mathrm{round}(\cdot)$ denotes the operation assigning a point in $\mathbb{R}^2$ to the nearest point in the hexagonal lattice $Q^{\mathrm{hex}}$, $r$ is a random vector with a decreasing magnitude (we will detail in \cref{ssec:setup}).

\begin{figure}[t]
  \centering
  \includegraphics[width=0.7\columnwidth]{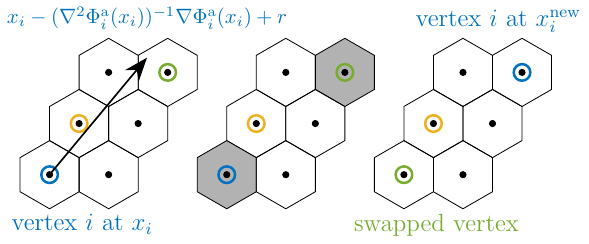}
  \caption{One iteration of the proposed algorithm.
    Step 1.
    Compute the coordinate Newton direction (blue).
    Step 2.
    Decide $x_{i}^{\mathrm{new}}$ by adding a random vector and rounding.
    Step 3.
    Move the vertex and swap the vertices if necessary (swap blue and green vertices).}
  \label{fig:hex}
\end{figure}

If there is a vertex $j$ such that $x_{j}= x_{i}^{\mathrm{new}}$, we swap the positions $x_{i}$ and $x_{j}$ to satisfy the condition $x_{i}\neq x_{j}$.
Otherwise, we just update $x_{i}$ to $x_{i}^{\mathrm{new}}$.
Refer to \cref{fig:hex} for a visual explanation.
Repeating this procedure yields an approximate solution to Problem~\eqref{eq:frApprox0}.

\subsection{Optimal Scaling}
\label{ssec:optimalScaling}

As the final step, we rescale the placement.
Let us formulate the optimization problem for the scaling factor $s > 0$.
For an initial placement $X = (x_{1}, \dots, x_{n})$, we scale it as $x_{i}\gets s x_{i}$ for all $i$.
The problem of minimizing $\Phi(X)$ through scaling is equivalent to minimizing $\phi(s)$ defined by
\begin{align*}
  \phi(s) \defeq{} & \sum_{(i,j) \in E}\frac{a_{i,j}(s \norm{x_i - x_j})^{3}}{3k}- k^{2}\sum_{i < j}\log(s \norm{x_i - x_j}), \\
  \phi'(s) ={}     & \sum_{(i,j) \in E}\frac{a_{i,j}\norm{x_i - x_j}^{3}}{k}s^{2}- \frac{k^{2}n(n-1)}{2s}.
\end{align*}
The function $\phi(s)$ is convex, and the optimal scaling factor $s^{*}$ satisfies $\phi'(s^{*}) = 0$:
\begin{equation}
  \label{eq:scaling}
  s^{*} = \qty(\frac{k^{3}n(n-1)}{2 \sum_{(i,j) \in E}a_{i,j}\norm{x_i - x_j}^{3}})^{1/3}.
\end{equation}
This value can be computed in $\order{\abs{E}}$ complexity, completing the initial placement algorithm.

We summarize the proposed algorithm in \cref{alg:proposed}.

\begin{algorithm}[t]
  \caption{Proposed algorithm for the initial placement}
  \label{alg:proposed} \KwIn{Graph $G = (V, E)$, Weight $(a_{i,j})_{(i,j) \in E}$, Parameter $N_{\mathrm{iter}}^{\mathsf{CN}}\in\bbN$.}
  \KwOut{Initial placement $X = (x_{1}, \dots, x_{n})$}
  Assign all vertices to distinct points in $Q^{\mathrm{hex}}$ randomly\;
  \For{$m \gets 0$ \KwTo $N_{\mathrm{iter}}^{\mathsf{CN}}$}{
  Select vertex $i \in V$ randomly\;
  $x_{i}^{\mathrm{new}}\gets \mathrm{round}(x_{i}- \nabla^{2}\Phi_{i}^{\mathrm{a}}(x_{i})^{-1}\nabla \Phi_{i}^{\mathrm{a}}(x_{i}) + r)$ with $r$ being a random vector\;
    \If{$\exists j \in V$ such that $x_{j}= x_{i}^{\mathrm{new}}$}{
      $x_{j}\gets x_{i}$\;
    }
    $x_{i}\gets x_{i}^{\mathrm{new}}$\;
    }
  $x_{i}\gets s^{*}x_{i}$ for all $i \in V$ with $s^{*}$ given by \cref{eq:scaling}\;
    \Return $X$
\end{algorithm}

\subsection{Application to Other Force Models}
\label{sec:otherForceModels}

As the end of this section, we state the application of our proposed method to other force models.
Although we have focused only on the FR force model so far, the concept of our approach can be applied to other models, as long as the objective function has a similar structure to $\Phi_{\mathrm{FR}}(X)$.
Recall that the energy minimization problem \eqref{prob:fr} is
\begin{mini*}
  {X \in \bbR^{2 \times n}} {\Phi_{\mathrm{FR}}(X) = \sum_{(i,j)\in E} \frac{a_{i,j}\norm{x_{i} - x_{j}}^{3}}{3k} - \sum_{\substack{(i,j) \in V^2 \\ i < j}} k^2\log{(\norm{x_{i} - x_{j}} + \epsilon^{\mathrm{r}})},}
  {} {}
\end{mini*}
where $\epsilon^{\mathrm{r}}$ is a small constant to prevent divergence.
Similar objective functions arise from other force models, such as the HC (Hooke--Coulomb) \citep{atallahAlgorithmsTheoryComputation2009} and Eades \citep{eades1984heuristic} force models, which are as follows:
\begin{mini*}
  {X \in \bbR^{2 \times n}} {
    \Phi_{\mathrm{HC}}(X) \defeq
    \sum_{(i,j)\in E}
    \frac{k_1(\lVert x_{i} - x_{j} \rVert - a_{ij})^2}{2}
    +
    \sum_{\substack{(i,j) \in V^2 \\ i < j}}
    \frac{k_2}{\lVert x_{i} - x_{j} \rVert + \epsilon^{\mathrm{r}}},}{}{}
\end{mini*}
\begin{mini*}
  {X \in \bbR^{2 \times n}} {\Phi_{\mathrm{Eades}}(X) \defeq
    \sum_{(i,j)\in E}
    k_1 \lVert x_{i} - x_{j} \rVert
    \left(
    \log \frac{\lVert x_{i} - x_{j} \rVert}{a_{ij}} - 1
    \right)
    +
    \sum_{\substack{(i,j) \in V^2 \\ i < j}}
    \frac{k_2}{\lVert x_{i} - x_{j} \rVert + \epsilon^{\mathrm{r}}}.}{}{}
\end{mini*}
See also \citep{6183577}.
In general, these optimization problems are more broadly treated as ``objective functions arising from graphs.''~\citep{wright2022optimization}
Clearly, these objective functions have a similar structure to $\Phi_{\mathrm{FR}}(X)$, which consists of an attractive term that depends on the edges and a repulsive term that depends on all pairs of vertices.

Thus, our proposed approach can be applied to these force models as well.
In these models, however, the Hessian matrix for the coordinate Newton direction may be indefinite, so the resulting Newton direction is not necessarily a descent direction.
This issue can be addressed by standard remedies in the optimization literature, such as adding a damping term \citep{nocedal1999numerical}, using regularization to ensure positive definiteness \citep{uedaRegularizedNewtonMethod2014a}, or clipping the maximum step size.
Hence, the coordinate Newton direction can still be used in almost the same way as in the FR model.
Another difference is that this direction depends on the scaling of $Q^{\mathrm{hex}}$; we therefore numerically optimized the scaling factor $s$ and applied this scaling before computing the initial placement.

We emphasize that the proposed method is not suitable for all graph-derived objective functions. A representative example is the KK model~\citep{kamadaAlgorithmDrawingGeneral1989}, whose objective function is
\begin{mini*}
  {X \in \bbR^{2 \times n}} {\Phi_{\mathrm{KK}}(X) \coloneqq \sum_{\substack{(i,j) \in V^2 \\ i < j}}  \frac{(\lVert x_{i} - x_{j} \rVert - l_{ij})^2}{2 l_{ij}^2}.}
  {} {}
\end{mini*}
This objective differs from those considered above because it consists only of all-pairs terms and has no principal edge-based term.
Since our approach computes the Hessian matrix for edge-defined terms, it is not directly applicable to the KK model.
Although one could compute the coordinate Newton direction for this objective directly, its behavior is not necessarily favorable; see also \cref{sec:rationale}.

For the KK model, SGD-based methods have been proposed~\citep{zhengGraphDrawingStochastic2019,ahmedMulticriteriaScalableGraph2022a} and are likely more appropriate.
These methods repeatedly select a pair of vertices and optimize with respect to that pair, making them complementary to our method, which repeatedly selects a single vertex and optimizes with respect to that vertex.

\section{Numerical Experiment}
\label{sec:experiment}

In this section, we evaluate the proposed algorithm with various numerical experiments.

\subsection{Experimental Setup}
\label{ssec:setup}

We first summarize the experimental setup.
We conducted all numerical experiments in this section using \Cpp{17} compiled with g++ 13.3.0.
The experiments were run on a Dynabook CZ/MWS laptop equipped with a 13th Gen Intel Core i7-1360P processor and \SI{16}{\giga\byte} of RAM.
The software environment was Ubuntu 24.04 LTS under Windows Subsystem for Linux 2 on Windows 11 Home 64-bit.

For a fair comparison, we implemented the simulation-based algorithm in \Cpp{} based on NetworkX version 3.3~\citep{hagbergExploringNetworkStructure2008}, SciPy 1.14.1~\citep{2020SciPy-NMeth}, and utilized the \Cpp{} L-BFGS~\citep{qiuYixuanLBFGSpp2024,okazakiChokkanLiblbfgs2024} library for the L-BFGS algorithm.
We also referred to the open-source code of the hexagonal grid~\citep{patelHexagonalGrids2013} and Graphviz version 2.43.0~\citep{ellsonGraphvizOpenSource2002}.

We used the $3 \times 2$ algorithms.
As initial placements, we used
\begin{enumerate}[label=\textbullet, itemsep=0pt, parsep=0pt]
  \item random initialization (\textsf{RAND-}),

  \item the SA initialization obtained by Simulated Annealing (\textsf{SA-})~\citep{ghassemitoosiSimulatedAnnealingPreProcessing2016}, and

  \item the proposed initialization obtained with coordinate Newton direction (\textsf{CN-}).
\end{enumerate}
As algorithms to solve the energy minimization problem, we used
\begin{enumerate}[label=\textbullet, itemsep=0pt, parsep=0pt]
  \item the simulation-based algorithm (\textsf{SIM}), and

  \item L-BFGS algorithm (\textsf{LBFGS}).
\end{enumerate}
The key remark is that we slightly modified the original simulation-based algorithm (presented in \cref{alg:fr}) so that it automatically adjusts the stepsize and isolates the effect of the initialization from that of the stepsize.
Specifically, we applied a simple backtracking line search at the first iteration to suppress overshooting.
The maximum number of iterations of the simulation-based and L-BFGS algorithms are $N_{\mathrm{iter}}^{\mathsf{SIM}}= N_{\mathrm{iter}}^{\mathsf{LBFGS}}= 150$ for the FR force model, and $N_{\mathrm{iter}}^{\mathsf{SIM}}= N_{\mathrm{iter}}^{\mathsf{LBFGS}}= 300$ for the HC and Eades force models.

As parameters, we set the number of iterations $N_{\mathrm{iter}}^{\mathsf{CN}}= 2 \abs{V}^{3}/ \abs{E}$ for~\cref{alg:proposed}, and we also set the total number of iterations of Simulated Annealing (\textsf{SA}) in~\cref{ssec:preprocessing} to the same value.
For a random vector $r$ in \cref{alg:proposed}, the magnitude is set to $1-0.5\cdot m / N_{\mathrm{iter}}^{\mathsf{CN}}$ for the $m$-th iteration, and its direction is uniformly distributed in $[0, 2\pi)$.
Since \cref{alg:proposed} requires a computational time proportional to the degree of the selected vertex in each iteration, the expected computational time per iteration is $\order*{\abs{E}/\abs{V}}$.
Consequently, we can roughly estimate that the total computational time of the proposed algorithm is $\order*{\abs{V}^2}$, equivalent to a few iterations of the FR or L-BFGS algorithms.
All the codes are available at GitHub~\citep{ThisPaperGitHub}.

\subsection{Plots and Visualizations}
\label{ssec:exprDetail}

\begin{figure*}[p]
  \includegraphics[width=\columnwidth]{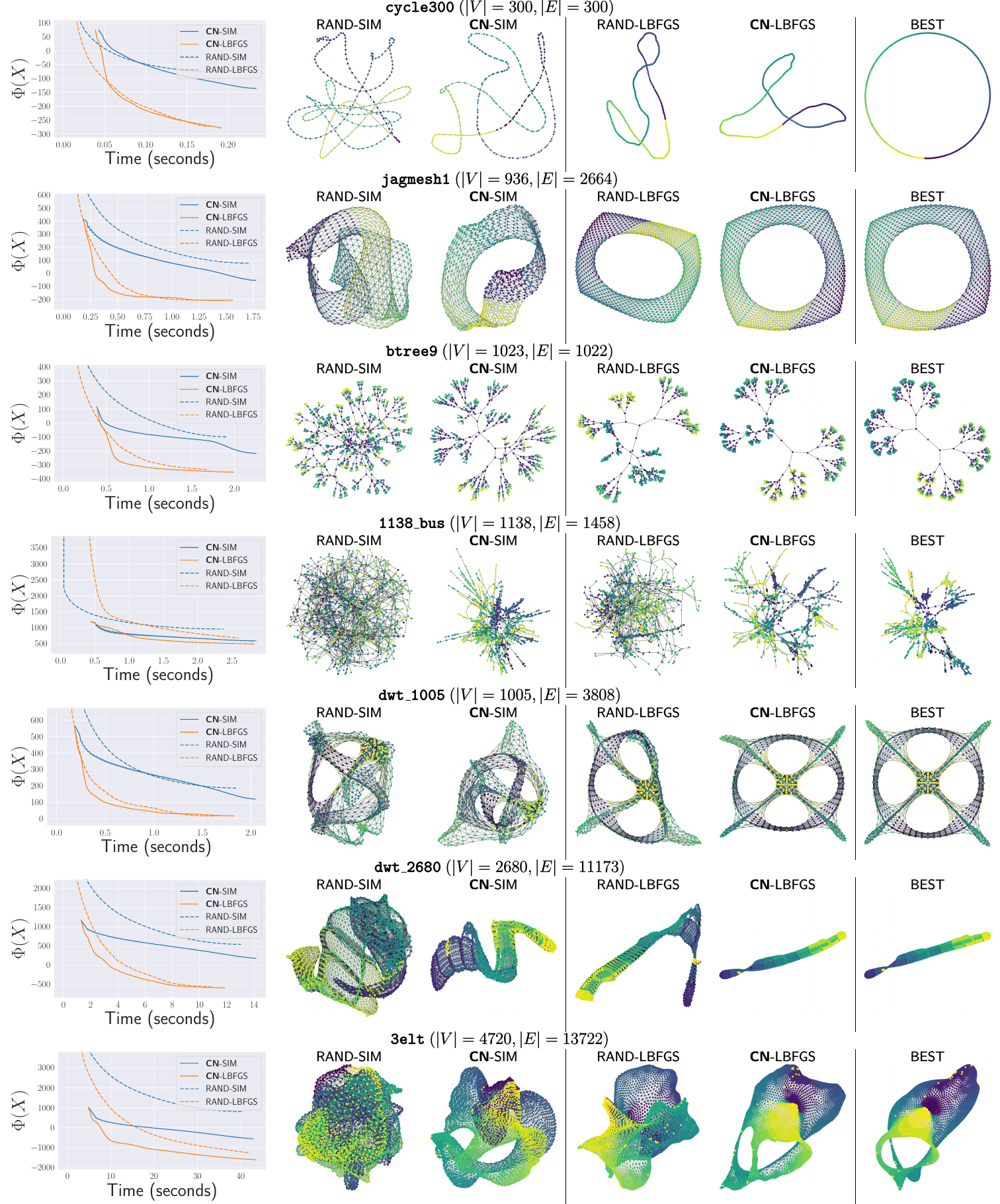}
  \makebox[\columnwidth][r]{%
    \makebox[0.8\columnwidth][c]{%
      \includegraphics[width=0.5\columnwidth]{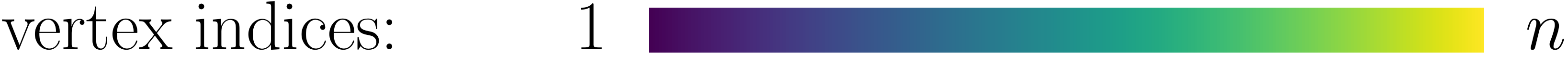}
    }%
  }
  \caption{
    Experimental results for the FR force model.
    Our proposed method \textsf{CN} yields improved results for both the simulation-based algorithm and L-BFGS.
  }
  \label{fig:individual}
\end{figure*}

We first show the plots and visualizations of the four algorithms in \cref{fig:individual}, \textsf{RAND-SIM}, \textsf{CN-SIM}, \textsf{RAND-LBFGS}, and \textsf{CN-LBFGS}.
We tested with 7 graphs: \texttt{cycle300}, \texttt{jagmesh1}, \texttt{dwt\_1005}, \texttt{btree9}, \texttt{1138\_bus}, \texttt{dwt\_2680}, and \texttt{3elt}.
Here \texttt{cycle300} is a cycle graph with 300 vertices, and \texttt{btree9} is a perfect binary tree with $2^{9+1}-1=1023$ vertices.
Other graphs are from the Sparse Matrix Collection~\citep{davisUniversityFloridaSparse2011}, and these choices are based on Ref.~\citep{zhengGraphDrawingStochastic2019}.
We provide more visualization results for various graphs in \cref{app:large_graphs}.

In \cref{fig:individual}, the plots on the left illustrate the objective function values $\Phi(X)$, the average of the five trials for each algorithm.
The graphs on the right are at the 100th iteration of the simulation-based and L-BFGS algorithms, and the graph at the bottom right is at the 1000th iteration of \textsf{CN-LBFGS} as a reference for the best solution.
The vertices in the graphs are colored according to vertex indices.

The observations and implications of \cref{fig:individual} are as follows.
First, the plots of the objective function values show that the proposed \textsf{CN} methods, represented by solid lines, start from substantially lower objective values.
This indicates that they produce high-quality approximate solutions within at most a few seconds, which is the main contribution of the proposed initialization.
Overall, the \textsf{CN} methods outperform their non-\textsf{CN} counterparts over the iterations.
In particular, \textsf{CN-SIM} consistently improves upon \textsf{RAND-SIM} across almost all graphs and iteration counts.
Similarly, \textsf{CN-LBFGS} often achieves lower objective values than \textsf{RAND-LBFGS}, although both sometimes converge to the same value.
This suggests that, given sufficiently many iterations, the effect of the initial layout on the final solution quality may diminish.
Under limited iteration budgets, the advantage of the proposed initialization remains substantial.

Second, the visualization results further support the effectiveness of the proposed initial placement.
In most cases, \textsf{CN} produces layouts that better approximate the nearly optimal arrangement, with more structured and less cluttered placements than random initialization.
Note that subtle differences in the objective values can still correspond to visible differences in the resulting layouts.
Additionally, the plot shows averages of the trials, whereas each visualization comes from a single trial.

As a side note, regardless of \textsf{CN} or non-\textsf{CN}, the algorithms with \textsf{LBFGS} consistently outperform those with \textsf{SIM}.
This finding is consistent with prior research~\citep{6183577}.

\subsection{Other Force Models}

\begin{figure*}[p]
  \includegraphics[width=\columnwidth]{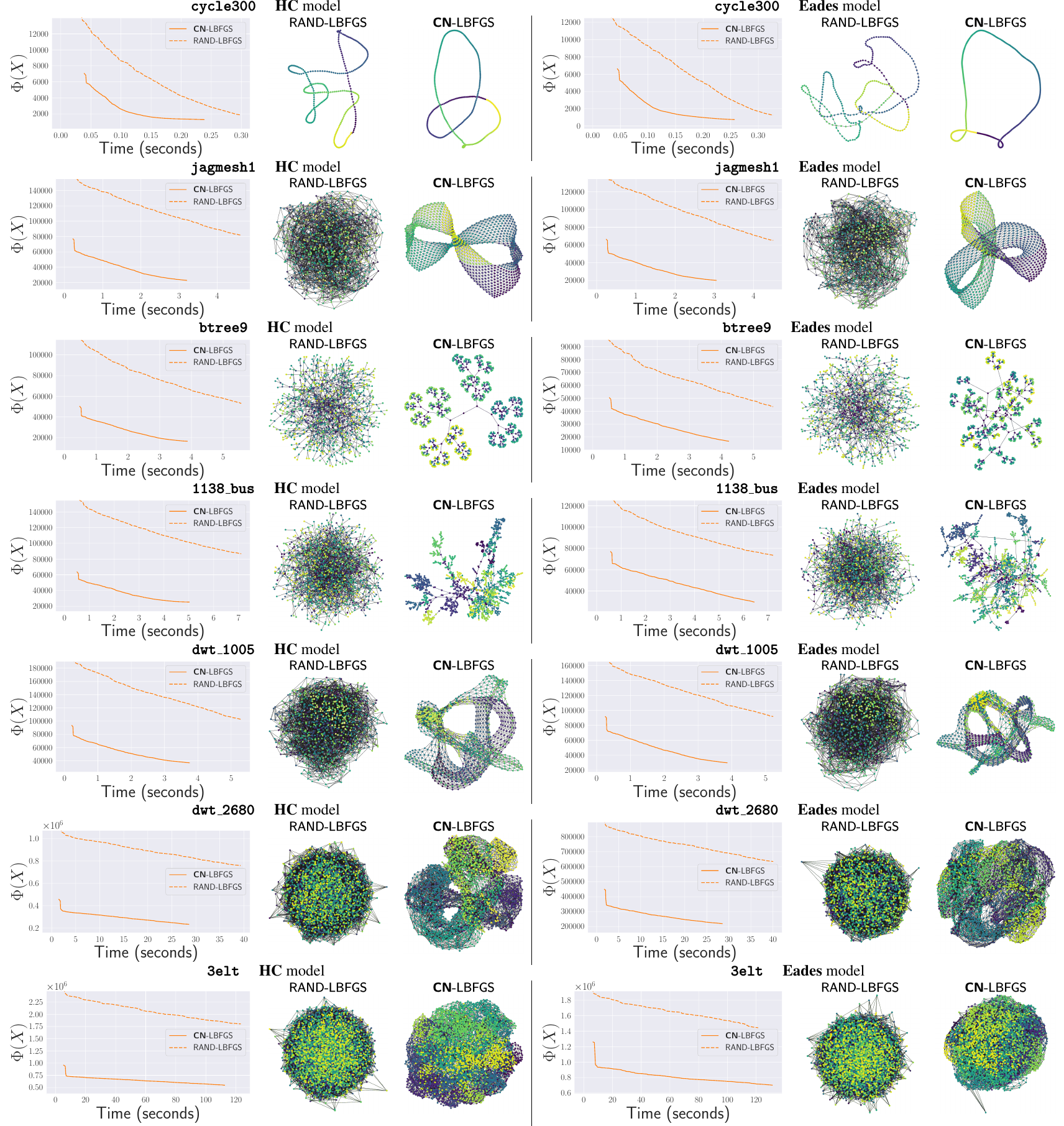}
  \makebox[\columnwidth][r]{%
    \makebox[\columnwidth][c]{%
      \includegraphics[width=0.5\columnwidth]{individual/legend.pdf}
    }%
  }
  \caption{
    Experimental results for the HC and Eades force models.
    Our proposed method \textsf{CN} yields improved results.
  }
  \label{fig:individual_HC_and_Eades}
\end{figure*}

As we have mentioned in \cref{sec:otherForceModels}, the proposed algorithm can be applied to other force models, such as the HC and Eades models.
The main differences from the experimental setup for the FR model are as follows.
(i) The graph weights are all set to one, since the models are quite sensitive to extreme edge weights.
(ii) The total number of iterations is 300 as mentioned in \cref{ssec:setup}.
(iii) We only run the L-BFGS algorithm since the simulation-based algorithm was originally designed for the FR model and thus is not expected to perform well for these models.

We showed the results for these models in \cref{fig:individual_HC_and_Eades}.
The proposed algorithm also performed better than random initialization for these models, and the differences in the graphs are quite expressive in some cases.
This is because these force models are much more difficult to optimize than the FR model, and thus the quality of the initial placement has a more significant effect on the final solution.

\subsection{Comparison with Other Initializations} 
\label{ssec:exprAll}

Next, we conducted experiments to evaluate the performance of the proposed algorithm (\textsf{CN}) compared to random initialization (\textsf{RAND}) and SA initialization (\textsf{SA}) with various graphs.
We used all undirected, connected, non-negative weighted graphs with $1{,}000$ or fewer vertices, which can be generated using the matrices from the Sparse Matrix Collection~\citep{davisUniversityFloridaSparse2011} as adjacency matrices.
In total, we tested with 124 graphs.
For fairness, when we compare with \textsf{SA}, we set all the weights to one since the SA initialization was originally designed for unweighted graphs.

For the algorithms, we set $N_{\mathrm{iter}}^{\mathsf{SIM}} = N_{\mathrm{iter}}^{\mathsf{LBFGS}}= 50$, the same as the default parameter in NetworkX~\citep{hagbergExploringNetworkStructure2008}.
Only when we compare \textsf{CN} algorithms with \textsf{RAND} algorithms, we set $N_{\mathrm{iter}}^{\mathsf{SIM}}= N_{\mathrm{iter}}^{\mathsf{LBFGS}}= 45$ for the \textsf{CN} algorithms, since the preprocessing step is expected to take as long as a few iterations of FR or L-BFGS, as mentioned in \cref{ssec:setup}.

\begin{figure*}[t]
  \centering

  \begin{minipage}{0.475\columnwidth}
    \centering
    \includegraphics[width=\columnwidth]{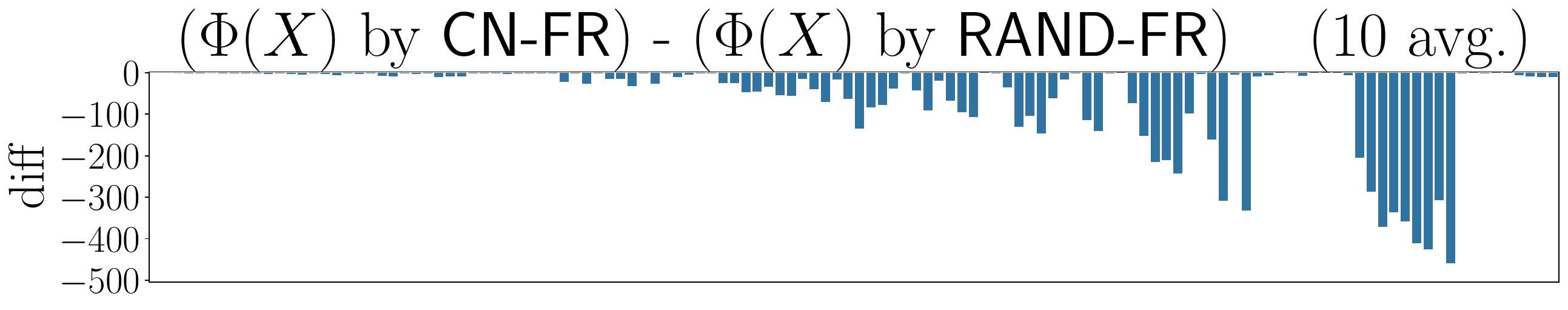}
  \end{minipage}
  \hfill
  \begin{minipage}{0.475\columnwidth}
    \centering
    \includegraphics[width=\columnwidth]{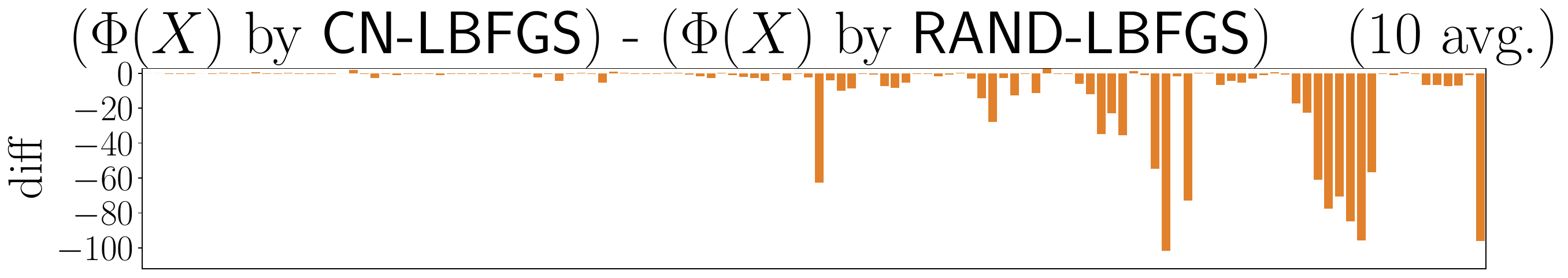}
  \end{minipage}

  \vspace{1em}

  \begin{minipage}{0.475\columnwidth}
    \centering
    \includegraphics[width=\columnwidth]{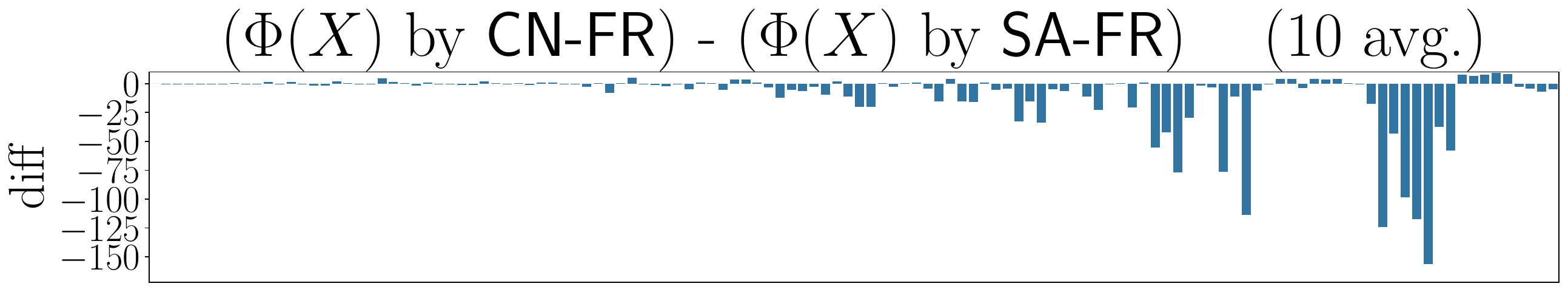}
  \end{minipage}
  \hfill
  \begin{minipage}{0.475\columnwidth}
    \centering
    \includegraphics[width=\columnwidth]{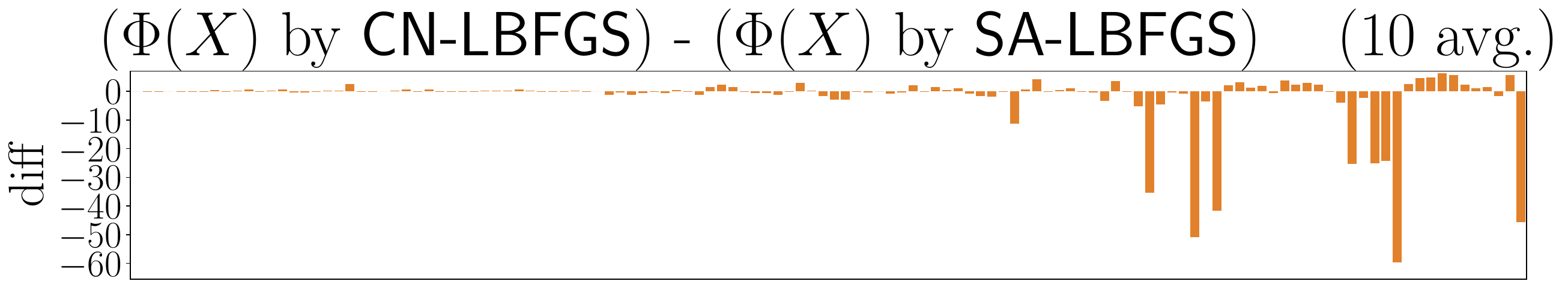}
  \end{minipage}

  \caption{
    Comparison of the initializations.
    The bins arranged along the x-axis correspond to the 124 graphs.
    The y-axis shows the difference in $\Phi(X)$ between \textsf{CN} and the baseline initialization method.
    The top panels compare \textsf{CN} with \textsf{RAND}, whereas the bottom panels compare \textsf{CN} with \textsf{SA}.
    Blue ones show results obtained using \textsf{SIM}, whereas orange ones show results obtained using \textsf{LBFGS}.
    In almost all cases for \textsf{RAND} and in most cases for \textsf{SA}, the differences are negative, meaning that the proposed initialization produced better final layouts than the baseline initialization methods.
  }
  \label{fig:overall-diff}
\end{figure*}

The results are shown in \cref{fig:overall-diff}.
The proposed algorithm performed better than random or SA initialization, except for a few cases.
These results strongly support the effectiveness of the proposed algorithm.
Even when the proposed algorithm performed worse, the difference was insignificant in almost all cases.

\begin{figure*}[t]
  \centering
  \begin{tabular}{ccccc}
    \toprule
                                                                                                             & \multicolumn{2}{c}{\texttt{dwt\_992} (\textsf{SA} better)}                           & \multicolumn{2}{c}{\texttt{collins\_15NN} (\textsf{CN} better)}                                                                                                                                                                                                            \\
                                                                                                             & initial placement                                                                    & 50th iteration                                                                      & initial placement                                                                         & 50th iteration                                                                           \\
    \cmidrule(lr){2-3} \cmidrule(lr){4-5}
    \raisebox{1.4em}{\rotatebox{90}{\textsf{SA} (\citep{ghassemitoosiSimulatedAnnealingPreProcessing2016})}} & \includegraphics[width=0.175\columnwidth]{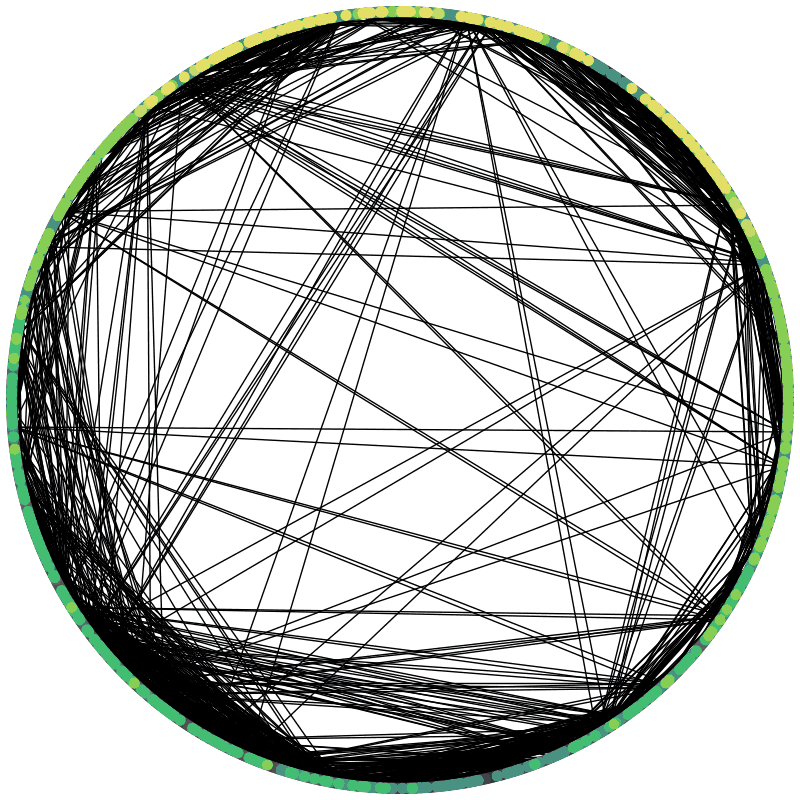} & \includegraphics[width=0.175\columnwidth]{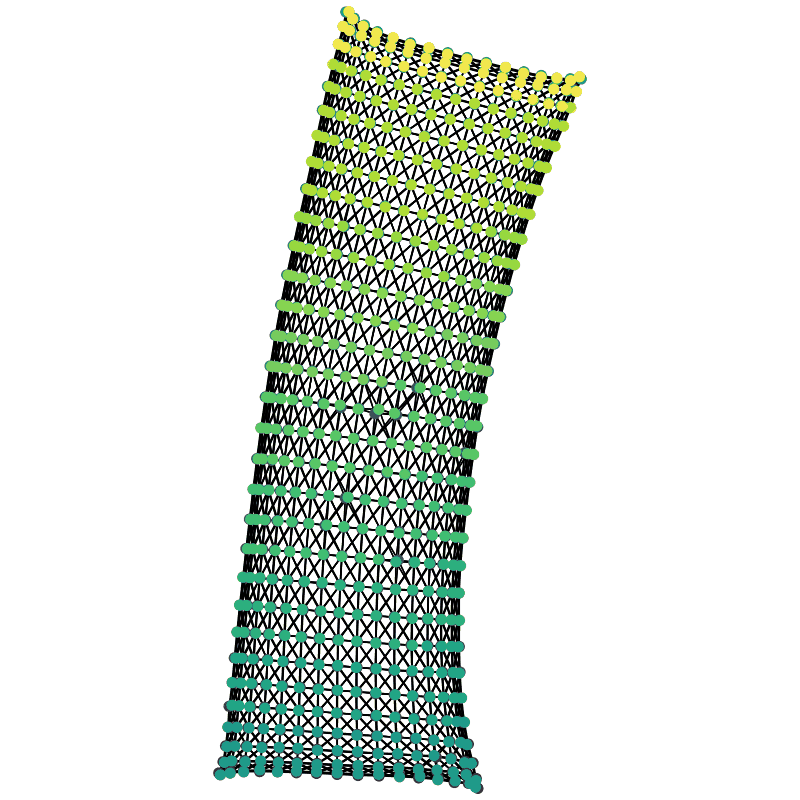} & \includegraphics[width=0.175\columnwidth]{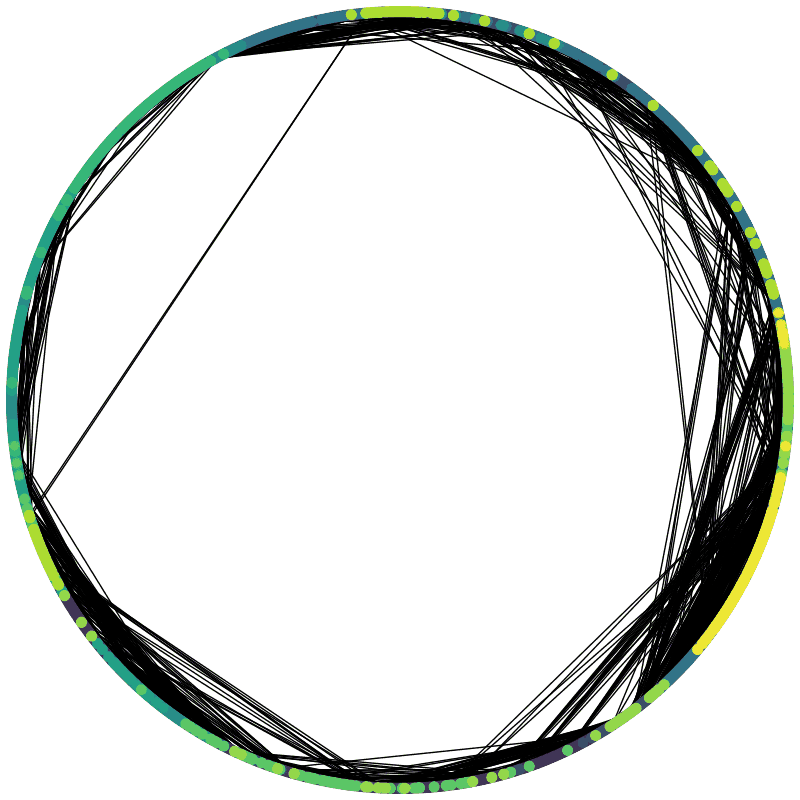} & \includegraphics[width=0.175\columnwidth]{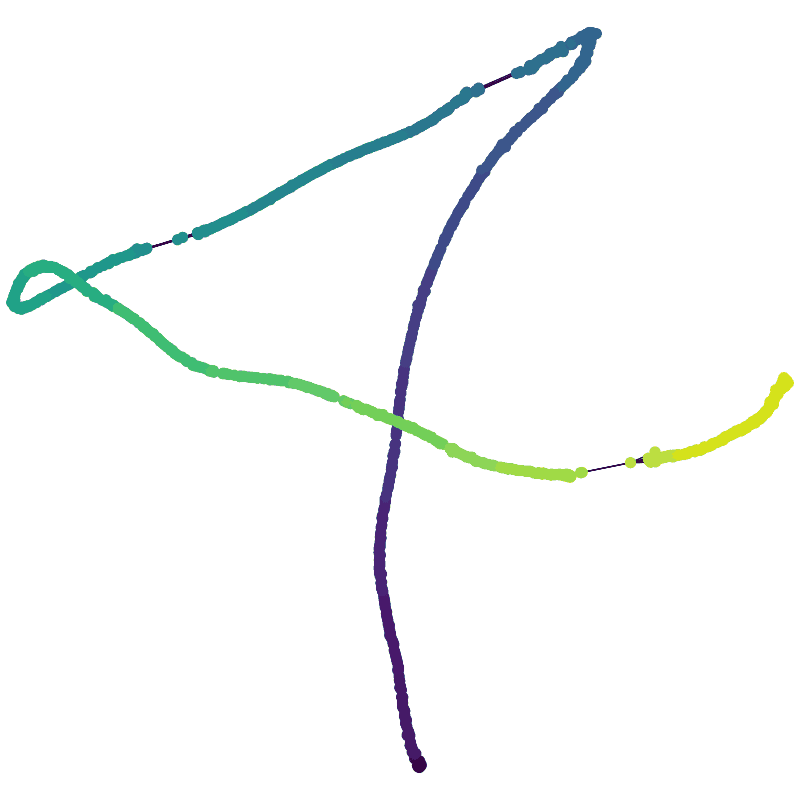} \\
    \raisebox{0.3em}{\rotatebox{90}{\textsf{CN} (proposed)}}                                                 & \includegraphics[width=0.175\columnwidth]{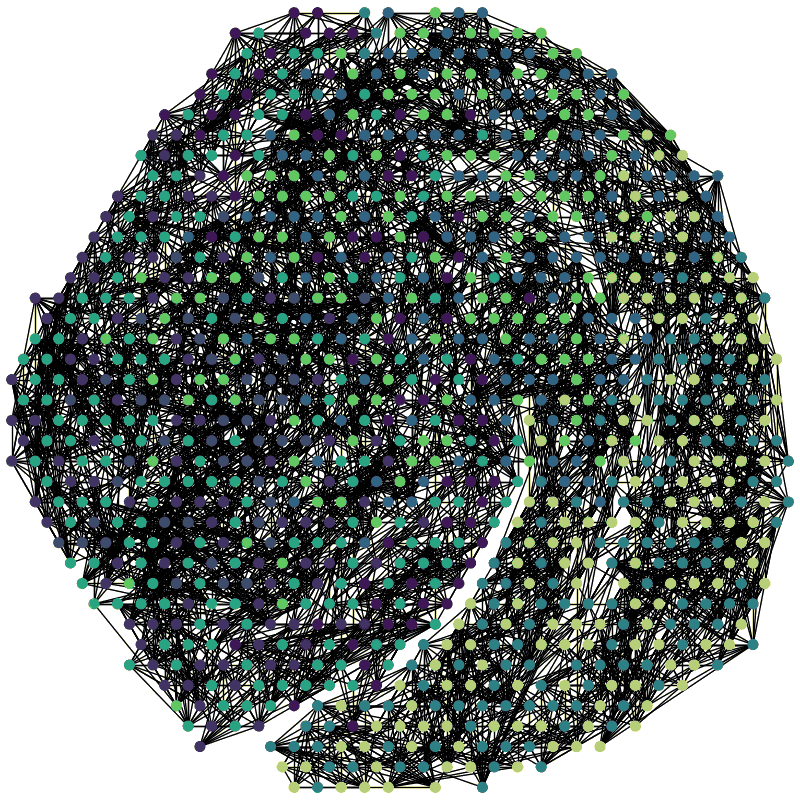} & \includegraphics[width=0.175\columnwidth]{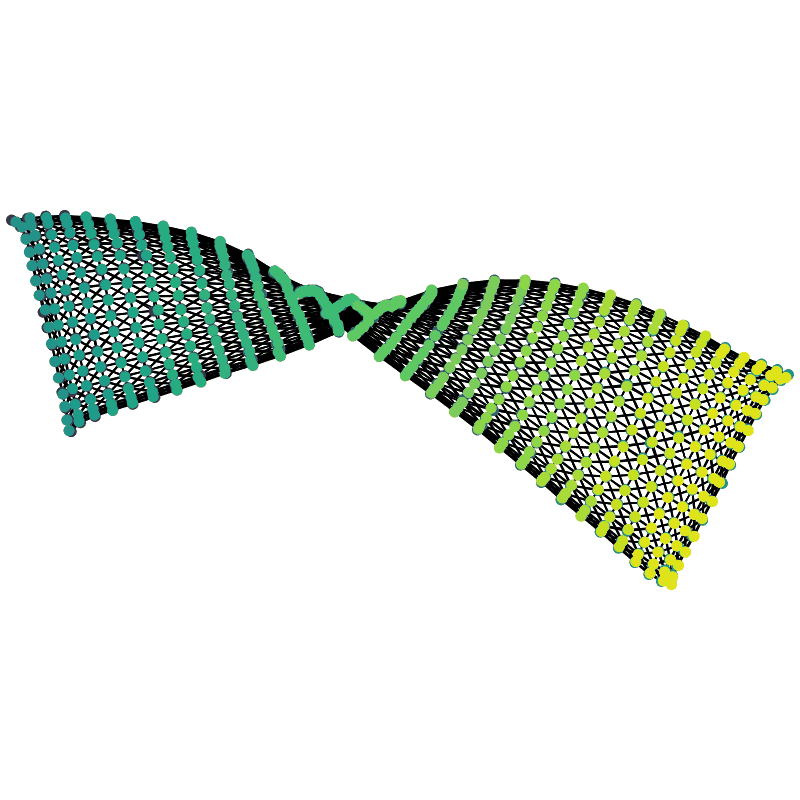} & \includegraphics[width=0.175\columnwidth]{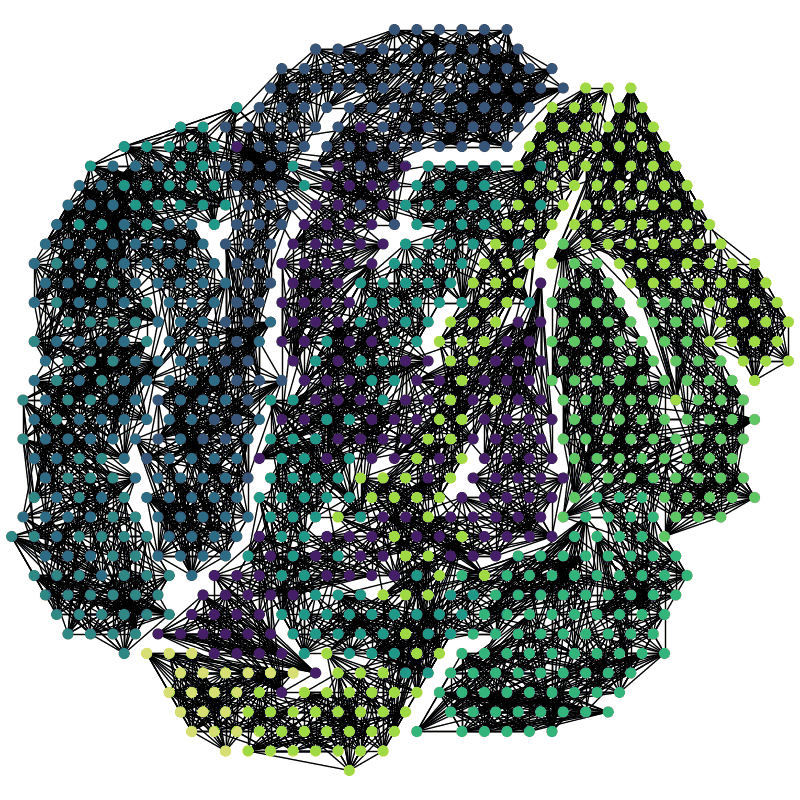} & \includegraphics[width=0.175\columnwidth]{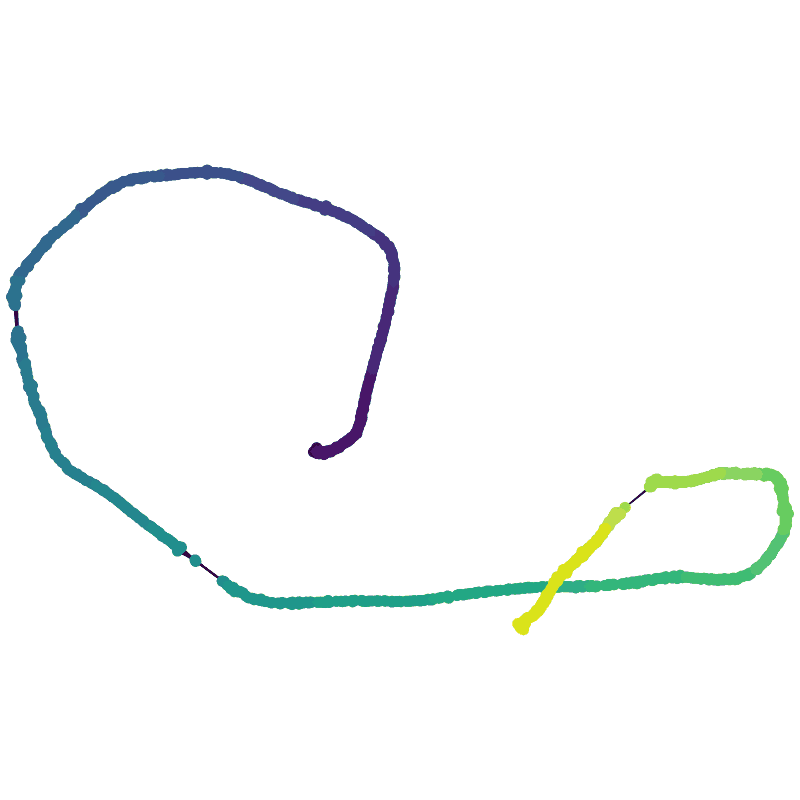} \\
    \bottomrule
  \end{tabular}
  \caption{Visualization results showing initial and 50th iteration placements for \texttt{dwt\_992} and \texttt{collins\_15NN}.}
  \label{fig:CN_vs_SA}
\end{figure*}

We also investigated cases in which the proposed algorithm performed worse than SA initialization in \cref{fig:CN_vs_SA}.
The proposed algorithm \textsf{CN} outperformed \textsf{SA} on \texttt{collins\_15NN}, whereas it underperformed on \texttt{dwt\_992}.

In many cases, the main factor that slows the convergence of subsequent simulation-based algorithms or L-BFGS algorithms is the presence of a twisted initial configuration, since untwisting such a configuration often requires many iterations.
Indeed, for \texttt{dwt\_992}, \textsf{CN} failed to resolve this twist, and the final layout remained twisted.
Conversely, for \texttt{collins\_15NN}, the initial layout produced by \textsf{SA} was twisted, causing the subsequent algorithm to spend additional time resolving the twist.

In general, it is difficult to predict how an initial layout will affect the final drawing.
This is because multiple factors interact in a complex manner, including the structure of the graph, the fixed positional constraints imposed by the initial layout, the characteristics of the optimization algorithm, and the random seed.

Although the proposed method uses a hexagonal lattice $Q^{\mathrm{hex}}$, alternative lattices or point sets, such as $Q^{\mathrm{circle}}$, may improve the performance of the proposed method in some cases.

\begin{figure*}[t]
  \centering
  \begin{minipage}{\columnwidth}
    \centering
    \begin{minipage}{0.4\columnwidth}
      \centering
      \includegraphics[height=2.8cm]{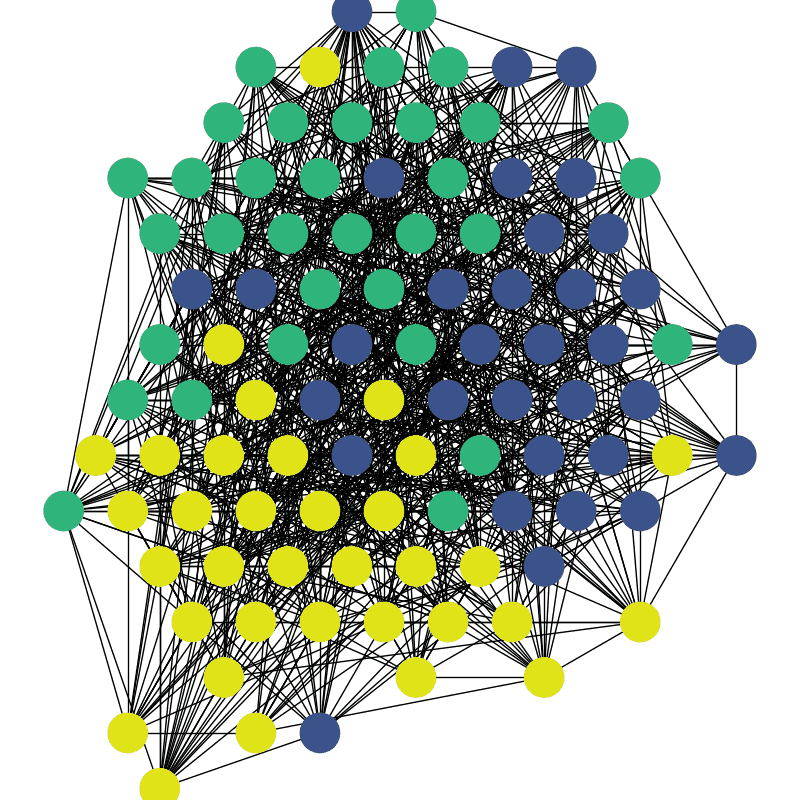}
    \end{minipage}
    \begin{minipage}{0.4\columnwidth}
      \centering
      \includegraphics[height=2.8cm]{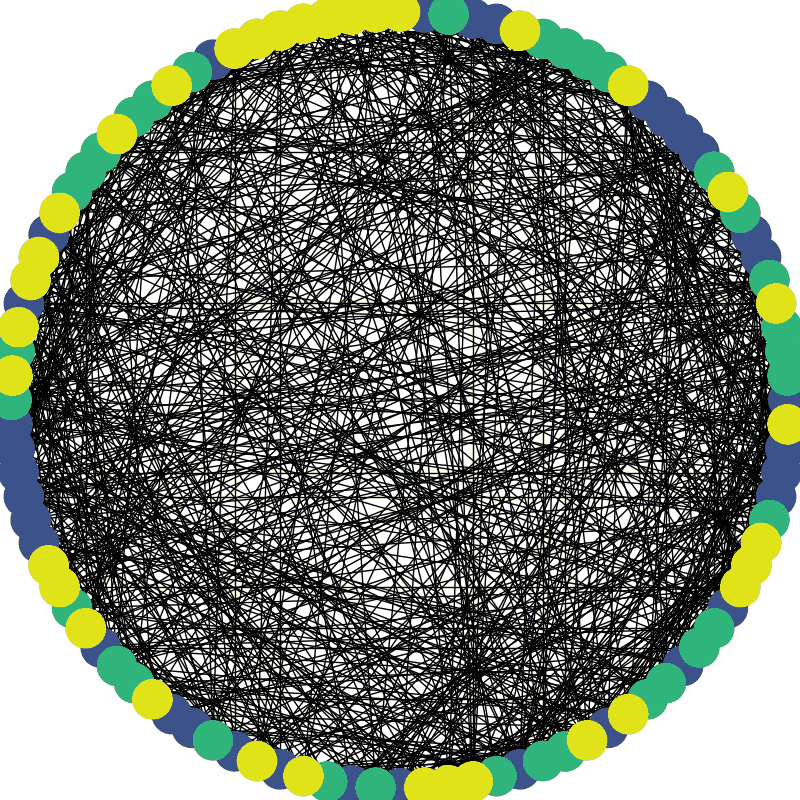}
    \end{minipage}
    \caption{
      An example with a weighted graph. The color indicates the group of vertices.
      Left: the result of \textsf{CN}, which is better as the vertices are separated by colors.
      Right: the result of \textsf{SA}, which is worse as there is no separation.
    }
    \label{fig:weightedDense}
  \end{minipage}
\end{figure*}

Furthermore, although this is a case not considered in Ref.~\citep{ghassemitoosiSimulatedAnnealingPreProcessing2016}, we also conducted experiments with \textsf{CN} and \textsf{SA} for a case where the graph is weighted, i.e., the edge weight $a_{i,j}$ is not necessarily in $\qty{0,1}$.
We generated a weighted graph with 100 vertices in three groups and 1000 edges randomly, and if the two vertices were in the same group, we set the edge weight to $1.0$; otherwise, we set it to $0.1$.
The difference between the two initializations is shown in \cref{fig:weightedDense}.
When we ignore the edge weights and just solve Problem~\eqref{eq:sa}, the graph is just an Erd\H{o}s--R\'enyi graph, and thus \textsf{SA} cannot find any meaningful structure in the initial placement.
In contrast, the proposed algorithm can identify the graph's structure, and we can observe that the left graph in \cref{fig:weightedDense} is separated into groups, as indicated by the node colors.
This result suggests that our proposed algorithm is effective even for weighted graphs, extending the applicability of the preprocessing step.

\section{Discussion}
\label{sec:discussion}

In this section, we discuss the implications of the proposed method and its potential applications and extensions.

\subsection{Application to Multilevel Approaches}
\label{sec:application_like_multilevel}

We briefly discuss how the proposed method can be combined with multilevel approaches.
Multilevel approaches, such as \textsf{sfdp}~\citep{Hu2006EfficientHF} in Graphviz, constitute important prior work for computing equilibrium layouts under force-directed models, that is, for solving the associated energy minimization problem.
Typically, a graph is recursively coarsened, laid out at the coarsest level, and refined through successive uncoarsening steps using simulation-based methods.

Our work is complementary to multilevel layout methods.
The proposed method could be used at the coarsest or selected intermediate levels as an alternative to the random initialization commonly used in multilevel approaches \citep{Hu2006EfficientHF,harelFastMultiScaleMethod2002}.
It could also be used during refinement: after a coarse vertex is expanded into finer vertices \citep{Hu2006EfficientHF,harelFastMultiScaleMethod2002}, the proposed initialization could be applied locally to the induced subgraph.

The proposed variants may also be combined with other refinement methods.
It is known that simulation-based refinement could be replaced by other methods \citep{walshawMultilevelAlgorithmForceDirected2003}, and our proposed placement or its variants can be used as an initial placement for such methods.
Although our implementation uses a hexagonal lattice, the essential requirement is the availability of sufficiently separated candidate positions.
Thus, similar initialization or coordinate-wise minimization strategies could be applied globally over the whole graph after coarsening.
Determining the most effective implementation, the appropriate level at which to apply the method, and the graph classes for which this integration is most beneficial remains an important direction for future work.
For further details on multilevel approaches, the reader is referred to~\citep{arleoMultilevelApproachEventBased2021}.

The proposed approach may also be useful for dynamic graph layouts, including animation and interactive visualization. Such scenarios are among the most prominent applications of the FR force model \citep{ertenGraphAELGraphAnimations2004,frishmanOnlineDynamicGraph2008}.
Since inexpensive updates are important in real-time large-graph settings, our algorithm's lower per-iteration cost than the standard FR algorithm may help initialize new vertices, improve local configurations, or accelerate refinement in scalable simulation-based dynamic layouts.
Since temporal coherence is central in dynamic graph drawing, integration with dynamic layout algorithms requires careful investigation, which we leave for future work.

\subsection{Rationale for the Discretization Approach}
\label{sec:rationale}

This section explains why we did not apply the coordinate Newton direction technique directly to the original Problem~\eqref{prob:fr}, but instead first transformed it into the discrete optimization Problem~\eqref{eq:frApprox0}.
At first sight, one might expect that applying the coordinate Newton method directly to~\eqref{prob:fr} would be more natural and possibly more efficient, since it avoids the detour through discretization.
However, as we shall see, this approach suffers from intrinsic difficulties that undermine its effectiveness.
In contrast, our discretization-based formulation allows us to circumvent these difficulties while still exploiting the advantages of the coordinate Newton direction.
This not only justifies our methodological choice but also provides a foundation for exploring further optimization strategies built upon this framework.
In what follows, we clarify the limitations of the direct approach and explain how our method addresses them.

\subsubsection{Possible Alternatives}
\label{ssec:possibleApproach}

One natural idea for solving the energy minimization problem is to apply stochastic coordinate descent, which resembles Randomized Subspace Newton methods~\citep{3454287.3454343, fujiTheoreticalAnalysisRandomized2025, cartisRandomisedSubspaceMethods2022, nozawaRandomizedSubspaceGradient2023, higuchiImprovingConvergenceGuarantees2024}.
For each vertex $i$, define the local energy
\begin{equation}\label{eq:fi}
  \Phi_{i}(x_{i}) \defeq \sum_{j \in V \setminus \{i\}} \Phi_{i,j}(\norm{x_i - x_j}),
\end{equation}
whose gradient (the net force on $i$) and Hessian are
\begin{gather*}\label{eq:gradientFi}
  \nabla \Phi_{i}(x_{i}) = \sum_{j \in V \setminus \{i\}} \qty(\frac{a_{i,j}\norm{x_i - x_j}}{k}- \frac{k^{2}}{\norm{x_i - x_j}^{2}}) (x_{i}-x_{j}),\\
  \nabla^{2}\Phi_{i}(x_{i}) = \sum_{j \in V \setminus \{i\}}\left(\frac{a_{i,j}\norm{x_i - x_j}}{k} - \frac{k^{2}}{\norm{x_i - x_j}^{2}}\right) \mqty(1&0\\0&1)+ \\
  \sum_{j \in V \setminus \{i\}}\left(\frac{a_{i,j}}{k \norm{x_i - x_j}}+ \frac{2k^{2}}{\norm{x_i - x_j}^{4}}
  \right) (x_{i}- x_{j})(x_{i}- x_{j})^{\top}.
\end{gather*}
A direct SCD scheme randomly selects a vertex $i$, applies Newton's method (or a regularized variant) to $\Phi_{i}$ using the above gradient and Hessian, and updates $x_{i}$ iteratively until convergence.

Although reasonable in spirit, this approach is also ineffective in practice.
In the following, we illustrate the reasons for this failure with nontrivial examples.

\begin{figure}[t]
  \centering
  \includegraphics[width=0.8\columnwidth]{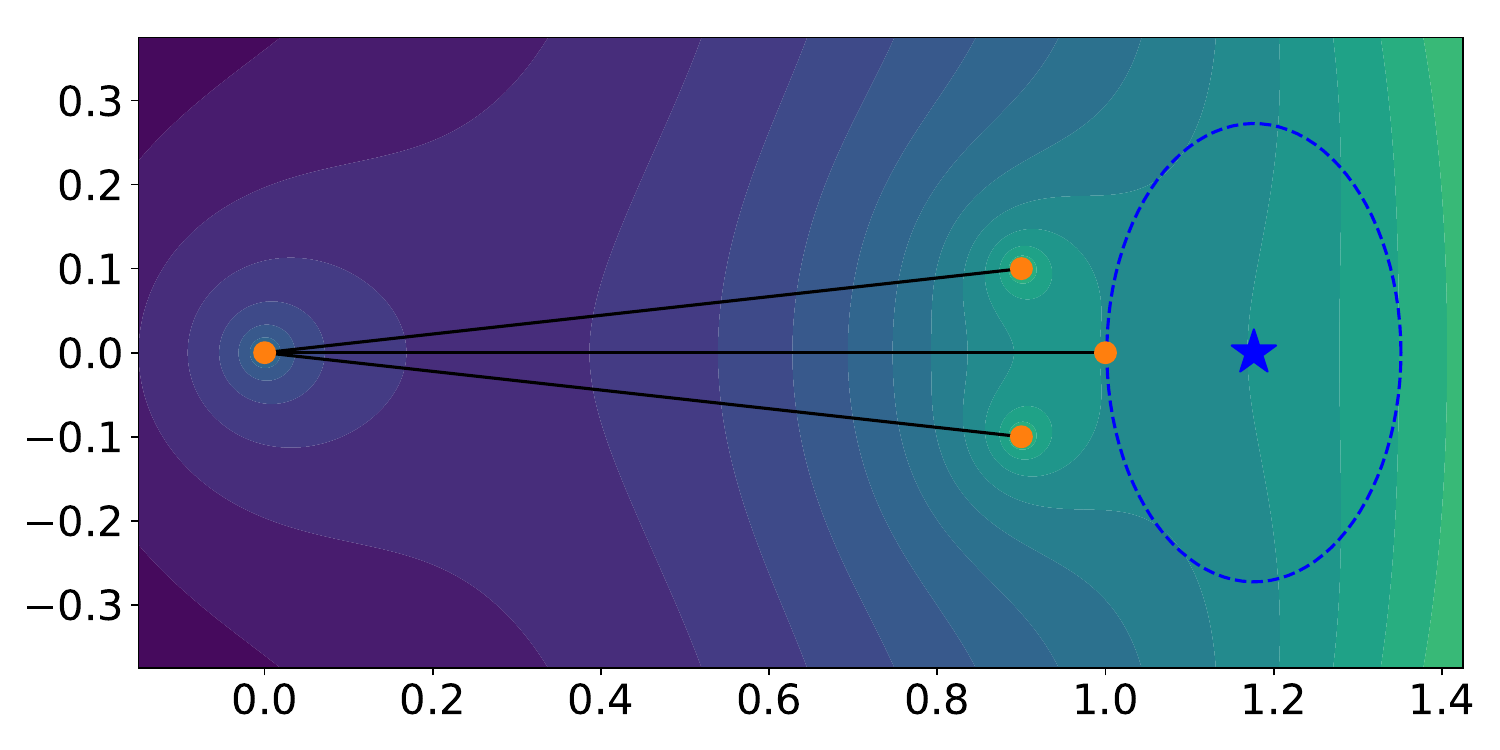}
  \caption{ The inaccurate quadratic approximation.
    The blue star indicates the optimal solution for the quadratic approximation of $\Phi_{1}(x_{1})$, but this differs significantly from the situation shown by the contour lines.
  }
  \label{fig:whyRSNfail}
\end{figure}

\subsubsection{Inaccuracy of Quadratic Approximation}
\label{ssec:inaccuracy}

One of the reasons it fails is the inaccuracy of the quadratic approximation; specifically, a particular issue arises when we restrict the optimization to a coordinate block.

Let $G$ be a graph with $V = \{1, 2, 3, 4\}$ and $E = \{(1, 2), (2, 3), (2, 4)\}$.
Set $k = 1/2$ and assign all positive edge weights $a_{i,j}= 1$ for every edge in $E$.
The position of the vertices is
\begin{equation*}
  X =
  \begin{pmatrix}
    1 & 0 & 0.9  & 0.9  \\
    0 & 0 & +0.1 & -0.1
  \end{pmatrix}.
\end{equation*}
We show the contour of $\Phi_{1}(x_{1})$ in \cref{fig:whyRSNfail}.
The key point of this example is that the Hessian
\begin{equation*}
  \nabla^{2}\Phi_{1}(x_{1}) = \mqty(4.25&0 \\
  0&1.75)
\end{equation*}
is positive definite and well-conditioned.
Despite this favorable property of the Hessian, the coordinate Newton direction for $x_{1}$ results in a deviation from the global optimum.
This issue arises from the inaccurate approximation of $\Phi_{1}$ in the restricted block coordinates $x_{1}$.
The attractive force from vertex $2$ and the repulsive forces from vertices $3$ and $4$ cancel each other out, leading to a highly inaccurate quadratic approximation.
This deficiency cannot be entirely resolved by modifying Newton's method, as it is an intrinsic and unavoidable limitation of stochastic coordinate descent.

\subsubsection{Ignorance of Other Vertices' Movements}
\label{ssec:ignorance}

Another reason is the ignorance of other vertices' movements when optimizing each vertex individually.
When optimizing for a vertex $i$, the coordinate Newton direction treats all other vertices $V \setminus \{i\}$ as fixed.

\begin{figure}[t]
  \centering
  \includegraphics[width=0.6\columnwidth]{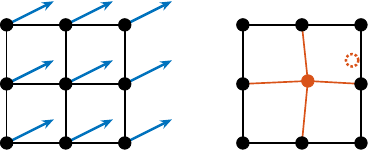}
  \caption{ The blue arrows represent the acting forces in this situation.
    The center vertex exhibits only a minor shift in the coordinate Newton direction, while the red dashed vertex marks its ideal configuration.
  }
  \label{fig:whyRSNfail2}
\end{figure}

\Cref{fig:whyRSNfail2} illustrates this issue.
Consider a subset of vertices forming a mesh-like structure in $G$, where all vertices receive forces in the directions indicated by the blue arrows.
In this situation, the FR and L-BFGS algorithms move all vertices simultaneously, allowing the simulation or optimization to proceed without issues.
In contrast, stochastic coordinate descent, as shown on the right side of the figure, brings stagnation.
Here, all other vertices are considered fixed.
Thus, optimizing the red vertex results in minimal movement, as its directly connected neighbors impede it.
As a result, even after numerous iterations, little optimization is achieved.
Thus, ignoring the movements of other vertices can be a significant limitation of the coordinate Newton method.

\subsubsection{Rationale for Proposed Method}
\label{ssec:rationale}

Our proposed method resolves issues in \Cref{ssec:inaccuracy,ssec:ignorance} by transforming Problem~\eqref{prob:fr} into Problem~\eqref{eq:frApprox0} in \cref{ssec:reduction}, and optimizing $\Phi^{\mathrm{a}}_{i}$ on $Q^{\mathrm{hex}}$.

Regarding the issue in \cref{ssec:inaccuracy}, this transformation brings the convexity of the objective function, as it treats the repulsive forces via a hexagonal lattice, making the quadratic approximation more accurate than the original problem.
Regarding the issue in \Cref{ssec:ignorance}, the discrete point set $Q$ ensures that the stepsize of any nonzero move is at least one, as each point is always separated by at least one.
This helps avoid very small coordinate updates and can improve exploration of the feasible set.
Furthermore, this transformation also brings the benefit of reducing computational complexity from $\order*{\abs{V}^2}$ to $\order{\abs{E}}$.

Thus, converting the problem is crucial to provide a high-quality initial placement with the coordinate Newton direction.
There may also be better transformation methods, and further exploration is warranted.

\section{Conclusion}\label{sec:conclusion}

In this study, we proposed a new initial placement with the coordinate Newton
direction for the FR force model, Problem~\eqref{prob:fr}. 
The obtained initial placements have fewer twists than the random initialization, leading to faster convergence and better visualization.
Numerical experiments revealed that the proposed method is effective across various graphs, extending the applicability of the preprocessing step.
The proposed method may advance graph drawing, and we also hope it highlights the potential of stochastic coordinate descent and its variants for addressing a broader range of graph-related optimization problems.

\section*{Acknowledgment}

We want to express our sincere gratitude to PL Poirion and Andi Han for their insightful discussions, which have greatly inspired and influenced this research.
We also thank the developers of NetworkX and Graphviz.
Their excellent work has been a great help in conducting this research.

This work was supported by JSPS KAKENHI Grant Numbers JP26KJ0936, JP23H03351, JP24K23853 and JST ERATO JPMJER1903.

\bibliographystyle{abbrvurl}
\bibliography{FruchtermanReingoldByRandomSubspace}

\appendix

\section{Additional Experimental Results}
\label{app:large_graphs}

\begin{figure*}[p]
  \includegraphics[width=\columnwidth]{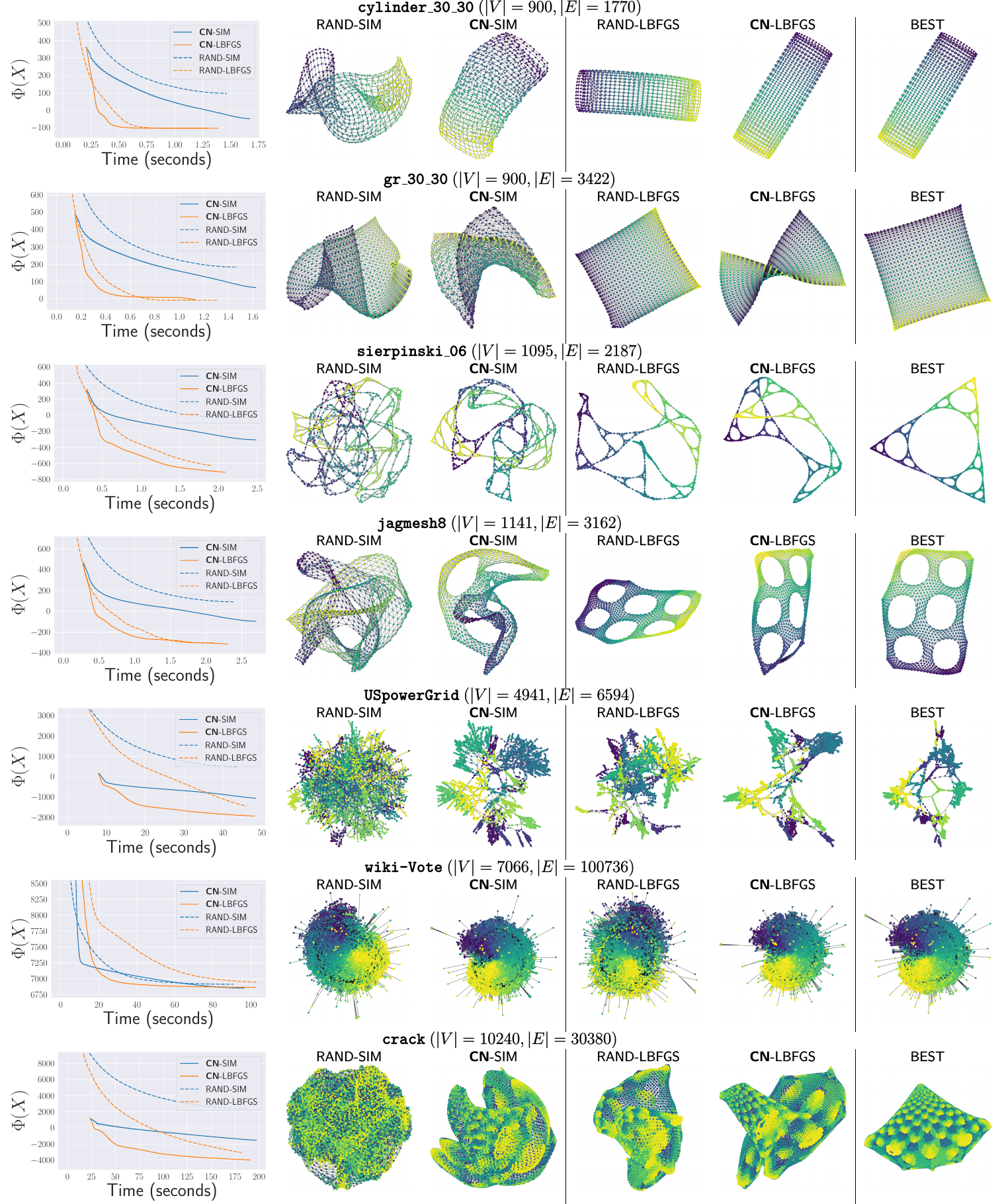}
  \makebox[\columnwidth][r]{%
    \makebox[0.8\columnwidth][c]{%
      \includegraphics[width=0.5\columnwidth]{individual/legend.pdf}
    }%
  }
  \caption{
    Additional experimental results for the FR force model.
    The objective function values at early states are generally improved for both simulation-based and L-BFGS algorithms, and the visual improvements by our proposed method \textsf{CN} are apparent for the simulation-based algorithms.
  }
  \label{fig:individual_2}
\end{figure*}

In addition to the evaluation in \cref{ssec:exprDetail}, we conducted experiments on several other graphs, with the results presented in \cref{fig:individual_2}.
For the choice of graphs, we referred to \citep{arleoDistributedMultilevelForceDirected2019,bartelExperimentalEvaluationMultilevel2011} and selected seven graphs: \texttt{cylinder\_30\_30}, \texttt{gr\_30\_30}, \texttt{sierpinski\_06}, \texttt{jagmesh8}, \texttt{USpowerGrid}, \texttt{wiki-Vote}, and \texttt{crack}.
The graph data were obtained from the SuiteSparse Matrix Collection~\citep{davisUniversityFloridaSparse2011} or generated by ourselves.

We can observe a similar trend to that in \cref{fig:individual}.
For the simulation-based algorithm, the effect of the proposed method is significant, and its visualization results well reflect the structure of the graph.
For L-BFGS algorithms, the positive effect is relatively smaller, but the objective function values in the plots are generally outperformed in most of the cases.
It is also worth noting that the visualization results shown on the right were produced using a single random seed.
They may appear worse than the expected performance shown in the plots.
Developing methods that are more robust to random seed selection remains a direction for future work.

As a side note, for larger graphs, the computational cost of convergence for both simulation-based algorithms and L-BFGS becomes substantially higher, and even a single iteration can require a considerable amount of time.
For further large-scale graphs with over ten thousand vertices, it will likely be necessary to combine the proposed method with a multilevel approach, as discussed in \cref{sec:application_like_multilevel}.

\section{NetworkX Implementation}
\label{sec:algToProb1}

As supplementary information, this section explains how to solve the energy minimization Problem~\eqref{prob:fr} and obtain the final placement.
We note that our implemented L-BFGS algorithm (without the proposed initialization) is available in NetworkX~\citep{hagbergExploringNetworkStructure2008}, and this section also serves as a reference for the details of the implementation in NetworkX.

\subsection{Simulation-Based Algorithm}
\label{ssec:frAlgorithm}

\begin{algorithm}[t]
  \caption{Simulation-Based (Fruchterman--Reingold) algorithm \citep{fruchtermanGraphDrawingForcedirected1991}}
  \label{alg:fr} \KwIn{Graph $G = (V, E)$, Weights $(a_{i,j})_{(i,j) \in E}$, Parameters $N_{\mathrm{iter}}^{\mathsf{SIM}}\in \bbN$, $t_{0}> 0$.}
  \KwOut{Final placement $X$}

  Initialize $X \in \bbR^{2 \times n}$\;
  $t \gets t_{0}$\;
  \For{$m \gets 1$ \KwTo $N_{\mathrm{iter}}^{\mathsf{SIM}}$}{
    $\text{compute gradient }\nabla \Phi_{i}(x_{i})$ for all $i \in V$\;
    $x_{i}^{\mathrm{new}}\gets x_{i}- t \frac{\nabla \Phi_{i}(x_{i})}{ \norm{\nabla \Phi_i(x_i)}}$ for all $i \in V$\;
    $x_{i}\gets x_{i}^{\mathrm{new}}$\;
    $t \gets t - t_{0}/ N_{\mathrm{iter}}^{\mathsf{SIM}}$\;
    \If{\text{convergence condition is satisfied}}{ \textbf{break}\; }
  }
  \Return $X$\;
\end{algorithm}

The simulation-based algorithm, known as the Fruchterman--Reingold algorithm~\citep{fruchtermanGraphDrawingForcedirected1991}, is the original and most standard approach to obtain the equilibrium state of the force model.
The pseudo-code is shown in \cref{alg:fr}, which is a variant of the gradient (steepest) descent method with the function $\Phi_{i}$ in \cref{eq:fi}~\citep{tunkelang1999numerical}.

\Cref{alg:fr} is based on the original pseudo-code~\citep{fruchtermanGraphDrawingForcedirected1991} and implementation in NetworkX~\citep{hagbergExploringNetworkStructure2008} with some omitted details.
We typically draw the initial placement $X$ from a uniform distribution on $[0,1]^{2 \times n}$.
The parameter $t$ denotes the temperature, which governs the stepsize along the steepest descent.
As the temperature gradually decreases, the algorithm converges to a particular placement, which is not necessarily a local optimum of Problem~\eqref{prob:fr}.

\subsection{L-BFGS Algorithm}
\label{ssec:lbfgs}

\begin{figure}[t]
  \centering
  \includegraphics[height=3.59cm]{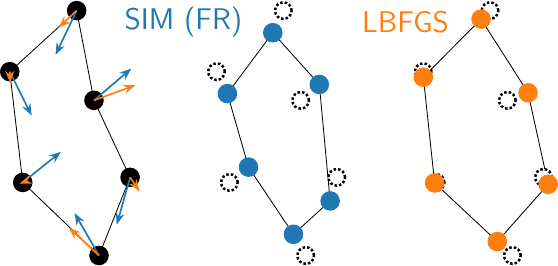}
  \caption{Comparison of the simulation-based and L-BFGS algorithms.
    Blue arrows represent fixed stepsizes in FR, whereas orange arrows represent adaptive stepsizes determined by the approximate inverse Hessian.}
  \label{fig:comparisonFRandLBFGS}
\end{figure}

Another approach is to leverage the L-BFGS algorithm~\citep{6183577}.
Using only a few recent gradient vectors, the L-BFGS algorithm approximates the inverse Hessian of the objective function $\Phi$~\citep{liuLimitedMemoryBFGS1989a}.
L-BFGS is known to be very efficient for large-scale optimization problems.
We can apply the L-BFGS algorithm by flattening the matrix $X \in \bbR^{2 \times n}$ to a vector $\overline{X}\in \bbR^{2n}$.
Refer to \cref{fig:comparisonFRandLBFGS} for a comparison to the simulation-based algorithm.

\subsection{NetworkX Implementation}
\label{subsec:networkX_impl}

\begin{figure}[t]
  \centering
  \includegraphics[width=\columnwidth]{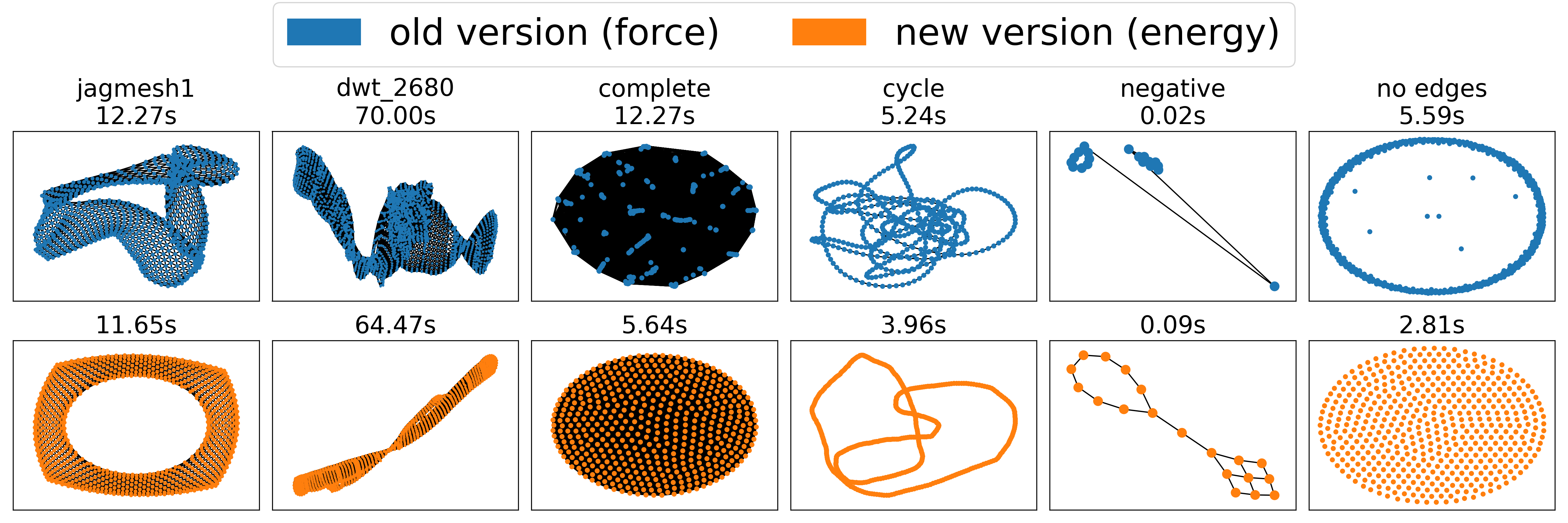}
  \caption{The simulation-based (FR) algorithm and L-BFGS algorithm in NetworkX.}
  \label{fig:networkx1}
\end{figure}

In this subsection, we explain the details of our NetworkX implementation.
Refer to \cref{fig:networkx1} for a comparison to the FR algorithm.
The problem setting considered in NetworkX is similar to that described in~\cref{sec:preliminaries}, but there are some key differences.
First, the number of dimensions is not always two.
Since the extension of the method is straightforward, we restrict our discussion to two-dimensional layouts.
Other than that, the following differences exist:
\begin{enumerate}[label=\textbullet, itemsep=0pt, parsep=0pt]
  \item Negative edge weights are allowed.

  \item Directed graphs are supported, not just undirected graphs.

  \item The graph is not always connected.
\end{enumerate}

Under these settings, the energy minimization problem~\eqref{prob:fr} can be unbounded.
It means that the optimal solution does not exist in a finite region.
When edge weights are negative, or the graph is disconnected, i.e., when it consists of $n_\mathrm{comp} >1$ connected components ($V=\bigcup_{j=1}^{n_\mathrm{comp}}V_{j}$), the solution may be unbounded, since the farther apart the vertices are, the lower the energy.

To overcome this issue, we do the following suitable modifications.
First, when the graph has negative edge weights, one may transform them into non-negative values, for example, by taking their absolute values.
Second, when the graph is a directed one, only the sum $a_{i, j}+ a_{j, i}$ matters in the formulation, so symmetrization is sufficient.
Thus, the directed and possibly negatively weighted adjacency matrix $\Tilde{A}= (\Tilde{a}_{i,j}) \in \mathbb{R}^{n \times n}$ can be transformed into a symmetric matrix $A = (a_{i,j})$, where $a_{i,j}= (\lvert \Tilde{a}_{i,j}\rvert + \lvert \Tilde{a}_{j,i}\rvert)/2 \geq 0$.
Third, we add additional forces to the energy function to prevent the divergence of each connected component.
In general, adding terms that attract each group to the center is a common approach to avoid unboundedness.
We adopt this approach with the center $x_{\mathrm{center}}=(0.5, 0.5)^{\top}$, as it is the center of the expected layout bounding box.

Based on \cref{sec:preliminaries}, let us define the sum of terms associated with the vertex $x_{i}$ as $\Phi_{i}$, which is explicitly given by
\begin{equation*}
  \Phi_{i}(x_{i}) \coloneqq \sum_{j \in V \setminus \{i\}}\left( \Phi_{i,j}(\norm{x_{i}- x_{j}}
    ) + \Phi_{j,i}(\norm{x_{j}- x_{i}}) \right).
\end{equation*}
The $i$-th row of the gradient $\nabla \Phi(X) \in \mathbb{R}^{2 \times n}$ is
\begin{align*}
  (\nabla \Phi(X))_{i}^{\top} & = \nabla \Phi_{i}(x_{i}) = \sum_{j \in V \setminus \{i\}}\left(\frac{(a_{i,j}+a_{j,i})\norm{ x_{i}- x_{j}}}{k}- \frac{2k^{2}}{\norm{ x_{i}- x_{j}}^{2}}\right) (x_{i}- x_{j}) \\
                              & = 2 \sum_{j \in V \setminus \{i\}}\left(\frac{a_{i,j}\norm{ x_{i}- x_{j}}}{k}- \frac{k^{2}}{\norm{ x_{i}- x_{j}}^{2}}\right) (x_{i}- x_{j}).
\end{align*}
To obtain the final layout while preventing the divergence, we need to minimize $\Phi(X) + f_{\mathrm{g}}(X)$, where $f_{\mathrm{g}}(X)$ is the additional term.
Let the center of gravity of the $j$-th connected component $V_{j}$ ($1 \leq j \leq m$) be $g_{j}\coloneqq \frac{1}{|V_{j}|}\sum_{i \in V_j}x_{i}$.
Then, using some constant $c_{\mathrm{g}}>0$, we define the additional force as
\begin{equation*}
  \Phi_{\mathrm{g}}(X) \coloneqq c_{\mathrm{g}}\sum_{j=1}^{m}\frac{\lvert V_{j}\rvert}{2}
  \norm{ g_{j} - x_{\mathrm{center}}}_{2}^{2},
\end{equation*}
with the gradient
\begin{equation*}
  (\nabla \Phi_{\mathrm{g}}(X))_{i}= c_{\mathrm{g}}(g_{j}- x_{\mathrm{center}}) \quad
  \text{where}\quad i \in V_{j}.
\end{equation*}
We can now apply the L-BFGS algorithm to these functions to derive the algorithm.
Further implementation details and experimental results are available in \href{https://github.com/networkx/networkx/pull/7889}{our merged pull request} to NetworkX.

\end{document}